 \documentclass[final,5p,times,10pt]{elsarticle}

 \usepackage{graphics}
\usepackage{graphicx}
\usepackage{url}
\usepackage{color}
\usepackage{bm}
\usepackage{array}
\usepackage{amsthm}

\biboptions{authoryear}

\newcommand{\grl}{Geophys.\ Res.\ Lett.}

\newcommand{\jgr}{Journal of Geophysical Research}

\newcommand{\ekin}{E_{\mbox{\scriptsize{kin}}}}
\newcommand{\cre}{cr}
\newcommand{\crf}{cr_f}

\usepackage{times}
\newcommand{\degree}{\ensuremath{^\circ}}
\journal{Earth and Planetary Science Letters}
\begin{document}
\begin{frontmatter}

\title{The 
Local-time variations of
Lunar Prospector epithermal-neutron data}

\author[lfat]{Lu\'{\a i}s~F.A.~Teodoro\corref{cor1}\fnref{fn1}}
\ead{luis.f.teodoro@nasa.gov}
\author[djl]{David J. Lawrence}
\author[vre]{Vincent R. Eke}
\author[rse]{Richard C. Elphic}
\author[wcf]{William C. Feldman}
\author[sm]{Sylvestre Maurice}
\author[mas]{Matthew A. Siegler}
\author[dap]{David A. Paige}
\cortext[cor1]{Lu\'{\a i}s~F.A.~Teodoro}

\address[lfat]{BAER Institute, Planetary Sciences Branch,  Space Sciences and Astrobiology Division, NASA Ames Research Center, Moffett Field, CA 94035-1000, USA}
\address[djl]{Mail Stop 200-W230, Johns Hopkins University Applied Physics Laboratory 11100 Johns Hopkins Road Laurel, MD 20723, USA}
\address[wcf]{Planetary Science Institute, 1700  E. Fort Lowell, Suite 106, Tucson, AZ, 85719, USA}
\address[vre]{Institute for Computational Cosmological, Department of Physics, Durham University, Durham, DH1 3LE, UK}
\address[rse]{Planetary Systems Branch, Space Sciences and Astrobiology Division, MS 245-3, NASA Ames Research Center,  Moffett Field, CA 94035-1000, USA}
\address[sm]{Universit\'{e} Paul Sabatier, Centre d'Etude Spatiale des Rayonnements, 9 avenue Colonel Roche, B.P. 44346 Toulouse, France}
\address[mas]{Planetary Science Institute, 1700  E. Fort Lowell, Suite 106, Tucson, AZ, 85719, USA}
\address[dap]{Department of Earth, Planetary, and Space Sciences, University of California, Los Angeles, CA 90095-1567, USA}

\begin{abstract}
We assess local-time variations  of epithermal-neutron count rates
 measured  by the Lunar Prospector Neutron Spectrometer. We investigate the nature 
of these  variations and find no evidence to support the idea that such variations are caused by diurnal variations of hydrogen concentration across the lunar surface.  Rather we find 
 an anticorrelation between instrumental  temperature and epithermal-neutron count rate. 
  We have also found that  the measured counts are dependent on the 
temperatures of the top decimeters of the lunar subsurface as constrained by the Lunar Reconnaissance Orbiter Diviner Lunar Radiometer Experiment temperature measurements. Finally, we have made the first measurement of the effective leakage depth for epithermal-neutrons of $\sim$20 cm.

 \end{abstract}

\end{frontmatter}

\section{Introduction}\label{sec:introduction}
The mapping of hydrogen on the near surface of the Moon
has been one of the focal points of the exploration of the Moon. 
Neutron spectroscopy has been central to this endeavor and allows one to
probe  the top few decimeters
 \citep[e.g.][]{Feldman:1998Sci...281.1496F,Lawrence:2006JGRE..11108001L}. 
Neutron spectrometers detect 
leakage neutrons resulting from the interaction of galactic cosmic rays (GCR) and the chemical elements present in the top layers of the planetary subsurface. These neutrons lose energy  by interacting with the surrounding lunar material, and different elastic and inelastic scattering cross-sections produce different amounts of moderation and capture. 
Since the Moon has no atmosphere,
cosmic rays reach the surface without being absorbed and
neutrons can leak from the near subsurface without being scattered
further or absorbed. 
In addition, the Moon has a characteristic composition pattern
(dichotomy between Fe and Ti-rich maria and Al-rich highlands),
to which neutron measurements are highly sensitive.
Finally, from returned samples, the lunar regolith
is recognized to be hydrogen poor at equatorial latitudes \citep[$<$100 ppm,][]{Heiken:1991}. This {\it a priori} information has played a 
major role in the interpretation and understanding of the orbital 
neutron data.

\medskip According to their kinetic energy, $\ekin$, neutrons are typically 
divided into three different energy bands: fast, epithermal and thermal. These correspond to $\ekin > 0.5\mbox{~MeV}$, $0.4\mbox{~eV}$ $<~\ekin~<~0.5\mbox{~MeV}$ and $\ekin \le~0.4\mbox{~eV}$ respectively. Briefly, thermal  neutrons are particularly sensitive to Fe, Ti, Gd and Sm in the absence of H \citep[][]{Elphic:1998Sci...281.1493E,Elphic:2000JGR...10520333E}.  
Epithermal-neutrons are strongly moderated by H \citep{Feldman:1993rgae.book.....P}
 and therefore provide a robust measure of H concentrations and their spatial variations.  Multiple studies have reported the presence and concentrations of H enhancements at the lunar poles \citep{Feldman:1998Sci...281.1496F,Eke:2009Icar..200...12E,Teodoro:2010GeoRL..3712201T,Teodoro:2014JGRE..119..574T}
as well as H concentrations within non-polar regions \citep{Lawrence:2015icarus}. 

\medskip In this article we study 
the daily variations of 
epithermal-neutron count rates from 
the Lunar Prospector neutron spectrometer (LPNS) to inquire if such potential variations are sensitive to daily variations in H concentrations.  This study is prompted, in part, by a report of time-variable concentrations of surficial (top tens of micrometers) H$_2$O/OH that have been observed with spectral reflectance data 
\citep{Sunshine:2009Sci...326..565S}.  If such time variable H concentrations extend to macroscopic depths of multiple mm to cm 
\citep{Lawrence:2011JGRE..116.1002L},
then such variations could in principle be observed with orbital neutron data.  
Preliminary analysis of data from the Lunar Exploration Neutron Detector (LEND) 
on the NASA Lunar Reconnaissance Orbiter (LRO) mission 
has shown a diurnal count rate variation, which has been interpreted as daily variations in H concentrations \citep{Livengood:2014LPI....45.1507L}. Such 
findings, if confirmed, could lead to important consequences to our 
understanding of the hydrogen distribution at the lunar surface.
Here, we will show that LPNS epithermal-neutron count rates also show  variations 
with solar local time, but that these variations are anti correlated with variations in instrumental temperature. Furthermore, 
the LPNS daily variations show 
a dependence on the temperatures at the top few decimeters of the lunar subsurface
as predicted by the numerical work of \cite{Little:2003JGRE..108.5046L} and \cite{Lawrence:2006JGRE..11108001L}.
These findings suggest that rather than reflecting diurnal changes in the hydrogen content of the lunar regolith, the temporal fluctuations in the count rates are due to small residual systematic effects in the data reduction.

\medskip In Section~\ref{sec:data_set} we present the datasets used throughout the article 
while in Section~\ref{sec:analysis} we perform the analysis of the epithermal dataset. We finish the article in Section~\ref{sec:conclusions}  with our conclusions and discussion.

\section{Datasets}\label{sec:data_set}
In this section we will briefly detail the main characteristics of the epithermal-neutron dataset 
gathered by the 
LPNS, which is
in the public domain at the {\it Planetary Data System} (PDS) (\url{http://pds-geosciences.wustl.edu/missions/lunarp/}).
We will also use temperature measurements of the lunar surface collected by the Diviner Lunar Radiometer
on Lunar Reconnaissance Orbiter. These  measurements are also available at PDS (\url{http://pds-geosciences.wustl.edu/missions/lro/diviner.htm}). 

\subsection{LPNS Counting data}\label{subset:counting_rate}
The Lunar Prospector Neutron Spectrometer,
designed and built at the Los Alamos National Laboratory,
consists of two gas proportional counters 
containing helium-3. 
 The neutrons that are captured by 
 helium-3 nuclei within the detectors 
  produce a unique energy signature that is then detected and counted. The two sensors 
 have different covers: 
one is covered in cadmium, the other in tin. The cadmium cover screens out thermal neutrons while the tin counterpart does not.  Differences in the counts between the two tubes thus quantify the detected number of thermal neutrons. 
The neutron spectrometer was located at the end of a 2.5 m boom, thus reducing the background from the main body of the spacecraft. For further details regarding the neutron spectrometer see \cite{Feldman:2004JGRE..109.7S06F}.
\begin{figure}[!t]
\begin{center}
\includegraphics[trim = 0.0mm 20.0mm 0.0mm 00.0mm, clip,width=0.95\columnwidth]{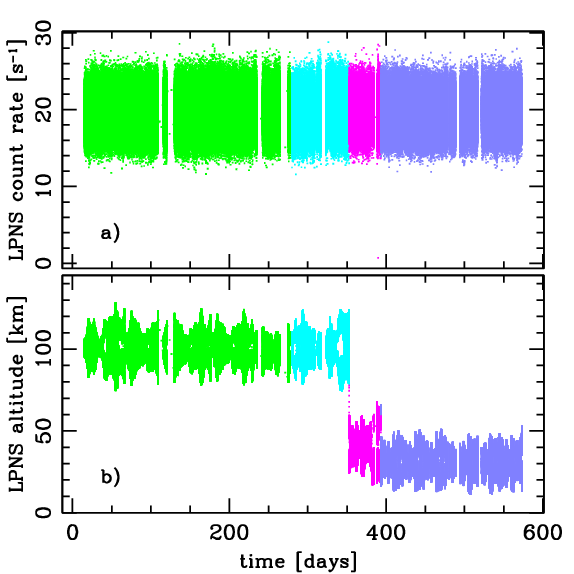}
\includegraphics[trim = 0.0mm 105.0mm 0.0mm 10.0mm, clip,width=0.95\columnwidth]{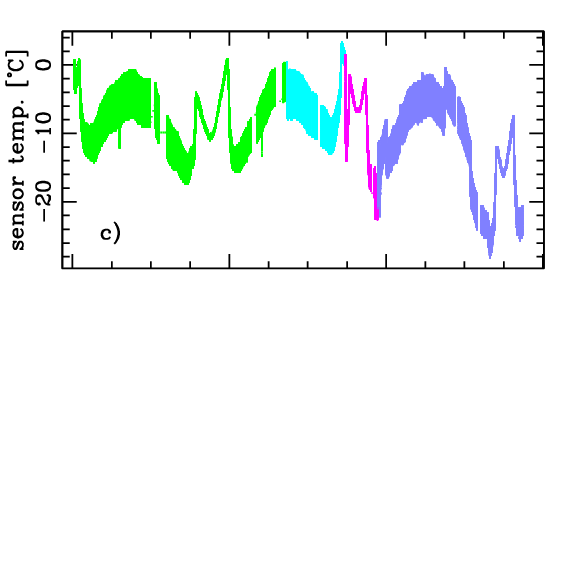}
\includegraphics[trim = 0.0mm 82.0mm 0.0mm 7.5mm, clip,width=0.95\columnwidth]{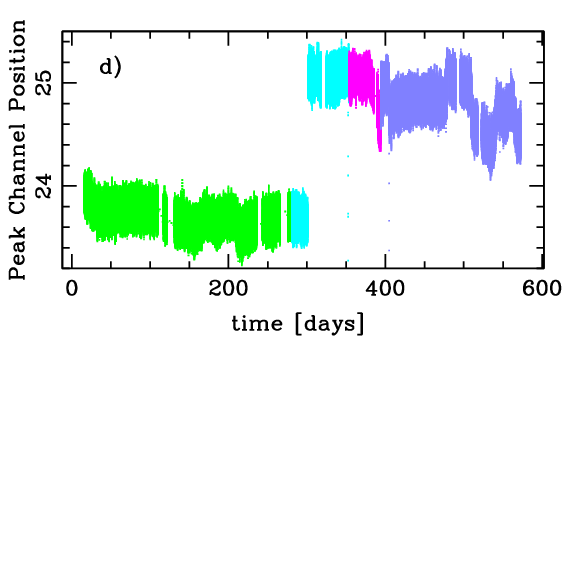}
\caption{ {\bf a)} Measured 8s counting rates  and {\bf b)} spacecraft altitude
versus time. We are only considering data in which $|\mbox{latitude}|<55^{\degree}$.
The colour scheme denotes different phases of the Lunar Prospector mission:
: High 1 (green), High 2 (cyan), magenta (Low 1), violet (Low 2). 
The origin of the global time was set at 1st of  January 1998. The individual counting rates were normalized to an altitude of 30 km \citep{Maurice:2004JGRE..109.7S04M}. 
{\bf c)}  Neutron sensor temperature. 
{\bf d)} Peak-channel position of the 764 keV neutron-capture peak in the LPNS pulse height spectra (see Figure~\ref{fig:peak_temp}).
The large jump in peak channel-position around time $\sim$300 days coincides 
with an increase in the sensor's high voltage.
}
\label{fig:lptimeseries}
\end{center}
\end{figure} 
\begin{figure*}[!t]
\begin{center}
\includegraphics[trim = 0.0mm 5.0mm 0.0mm -8.0mm, clip,height=0.52\columnwidth]{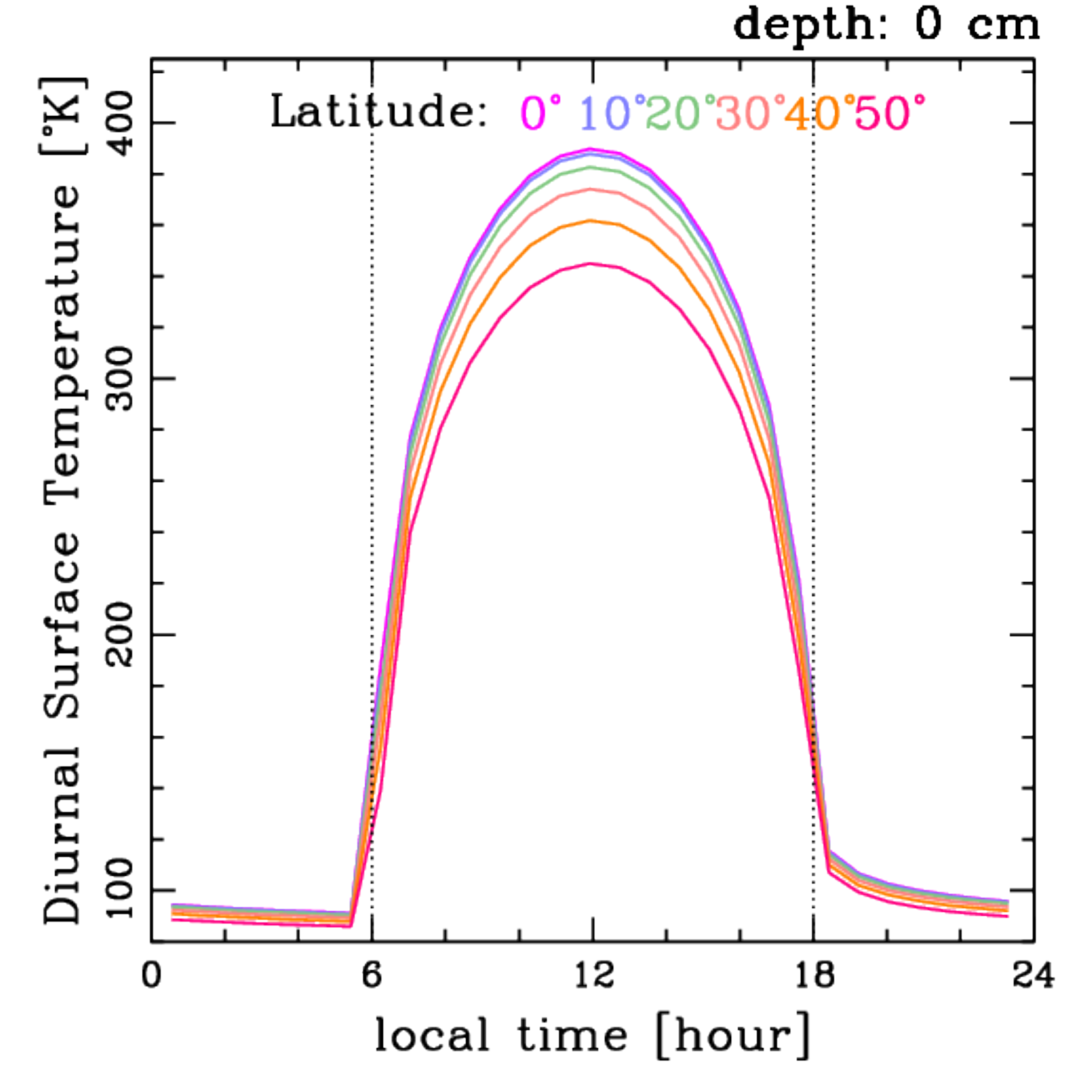}
\includegraphics[trim = 0.0mm 5.0mm 0.0mm -8.0mm, clip,height=0.52\columnwidth]{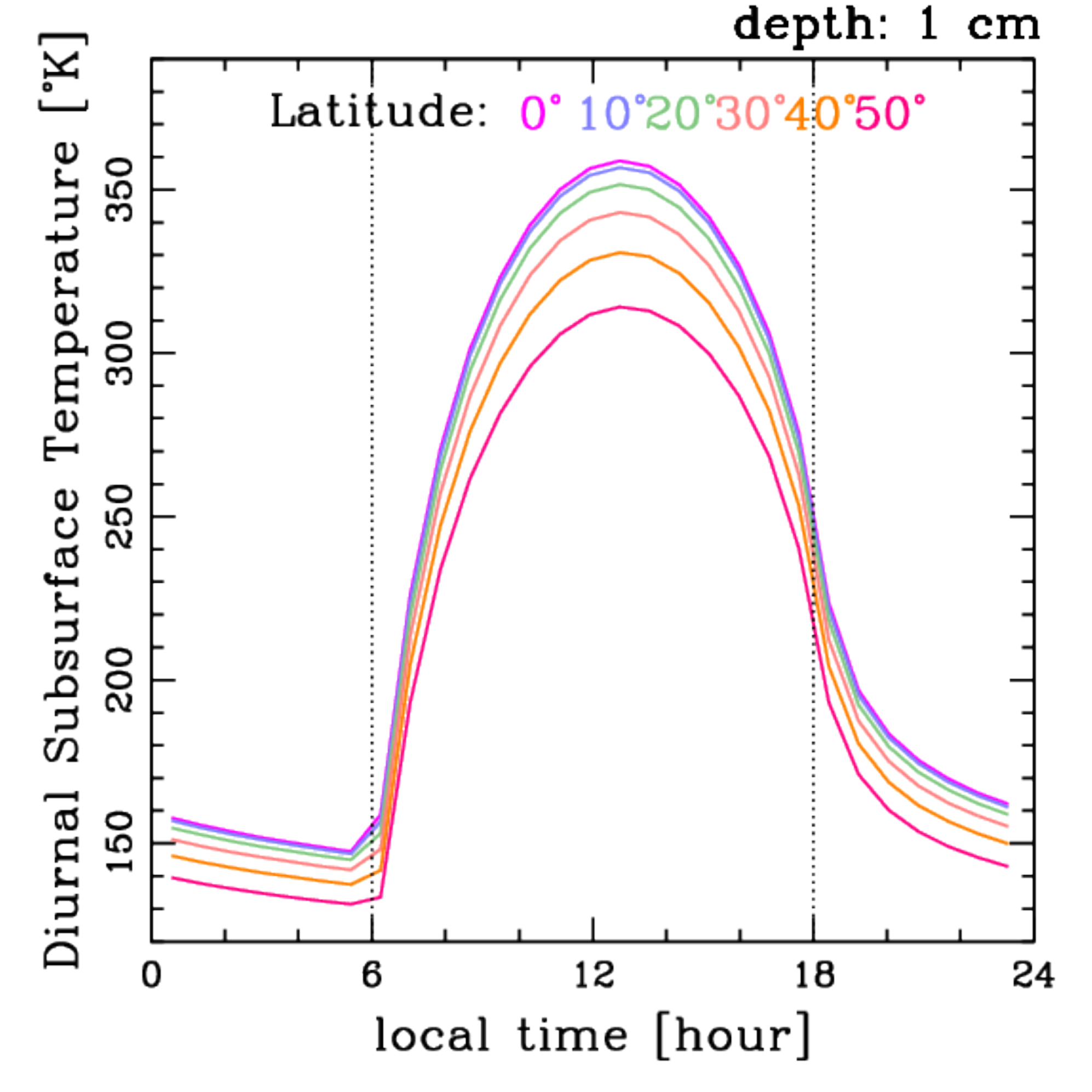}
\includegraphics[trim = 0.0mm 5.0mm 0.0mm -8.0mm, clip,height=0.52\columnwidth]{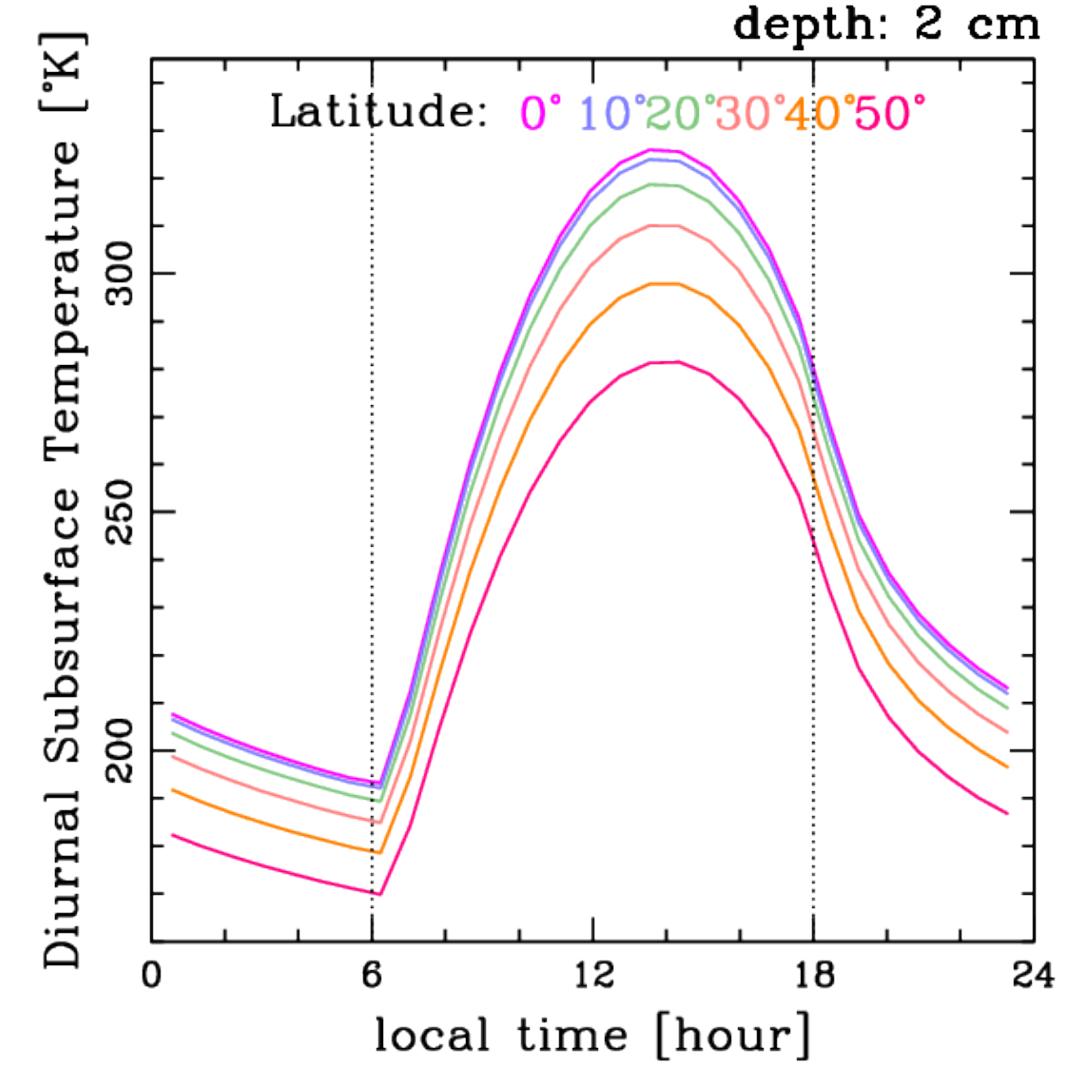}
\includegraphics[trim = 0.0mm 5.0mm 0.0mm -8.0mm, clip,height=0.52\columnwidth]{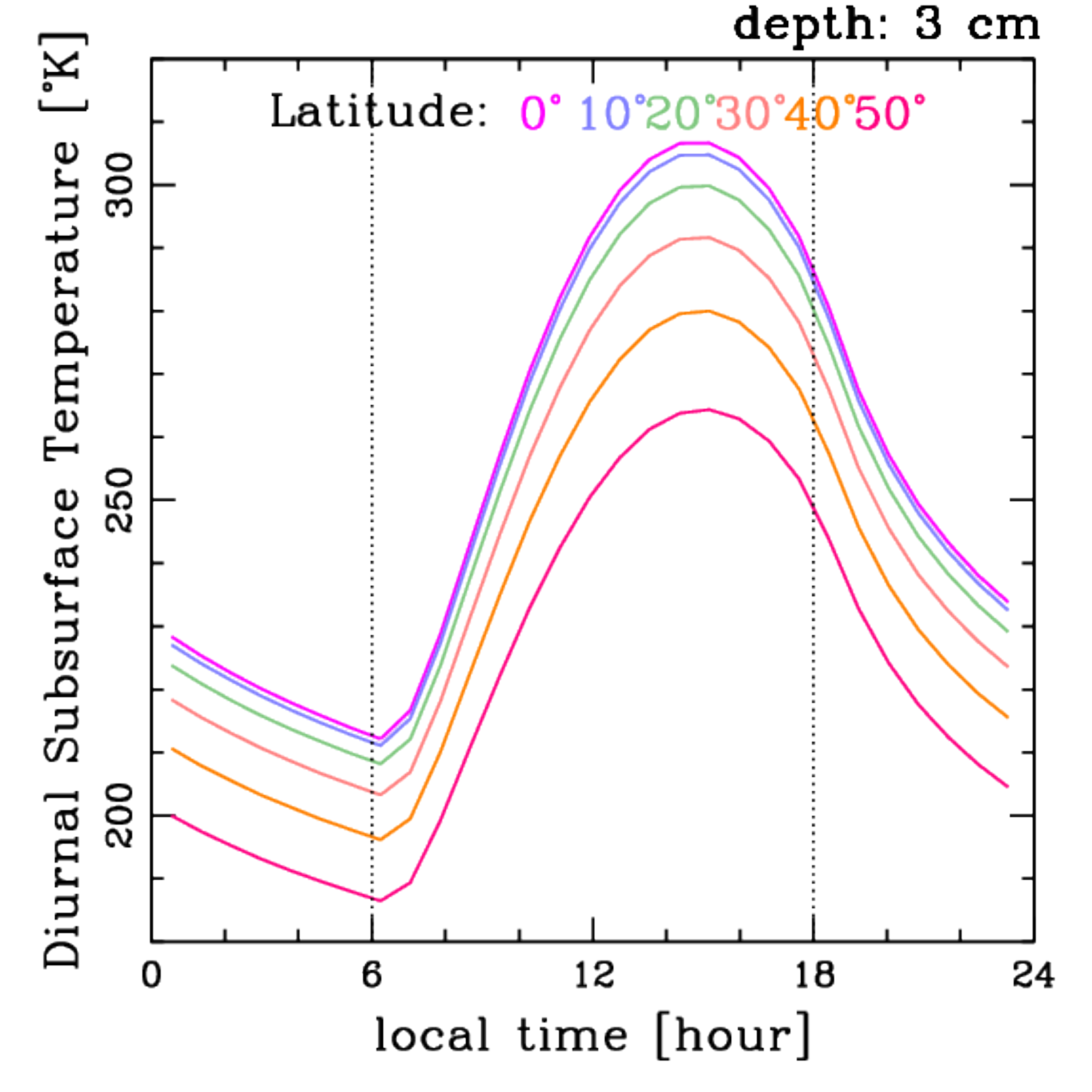}
\includegraphics[trim = 0.0mm 5.0mm 0.0mm -8.0mm, clip,height=0.52\columnwidth]{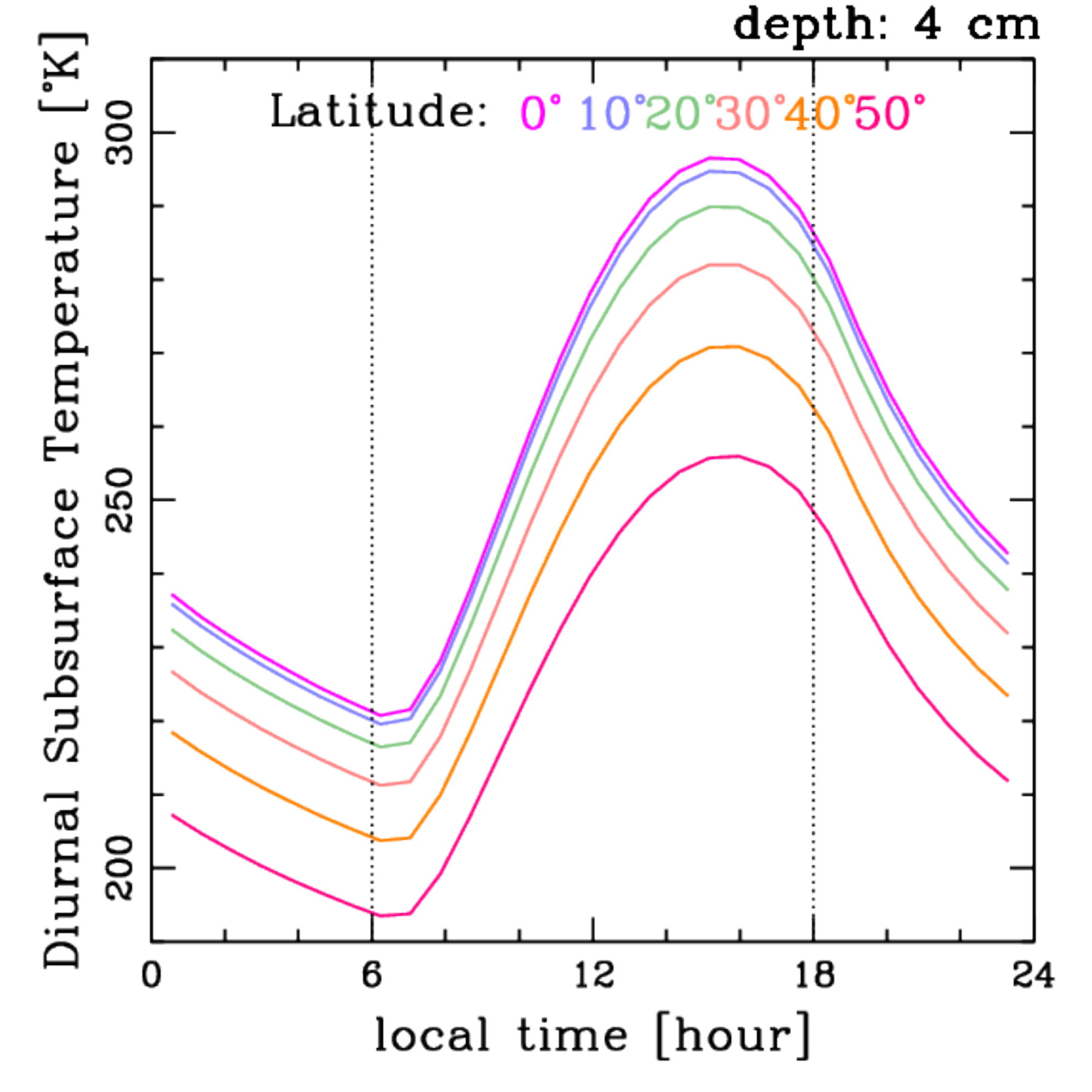}
\includegraphics[trim = 0.0mm 5.0mm 0.0mm -8.0mm, clip,height=0.52\columnwidth]{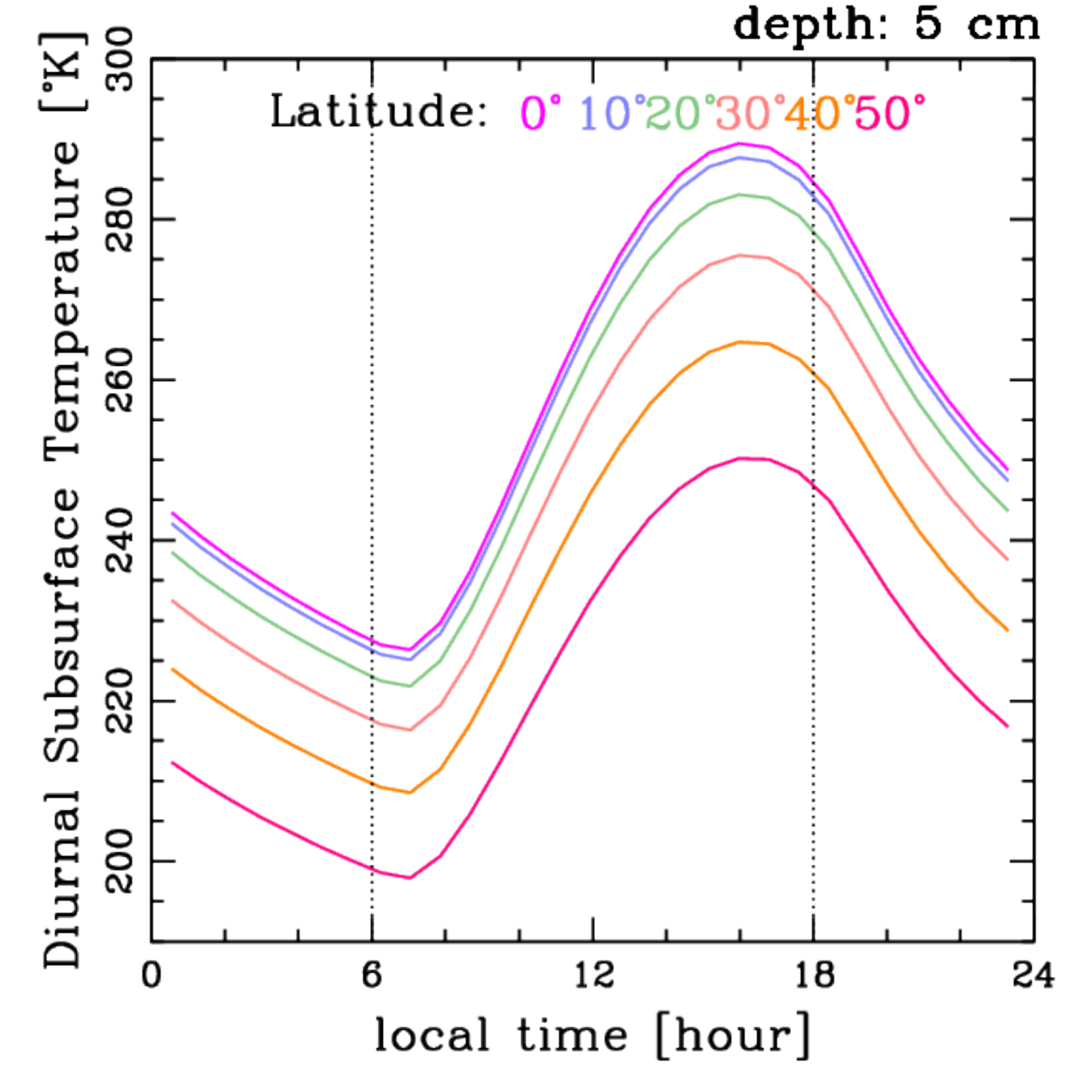}
\includegraphics[trim = 0.0mm 5.0mm 0.0mm -8.0mm, clip,height=0.52\columnwidth]{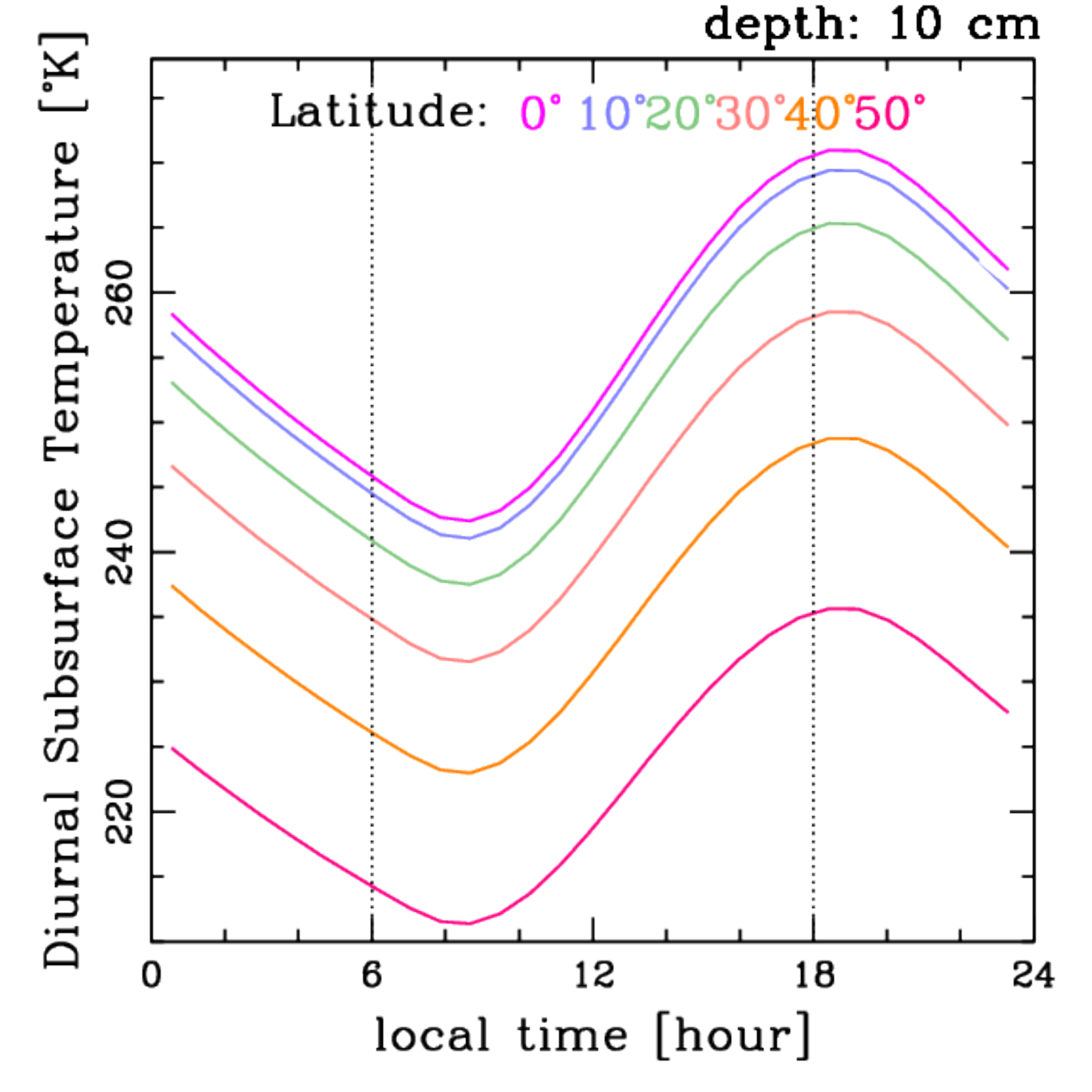}
\includegraphics[trim =0.0mm 5.0mm 0.0mm -8.0mm, clip,height=0.52\columnwidth]{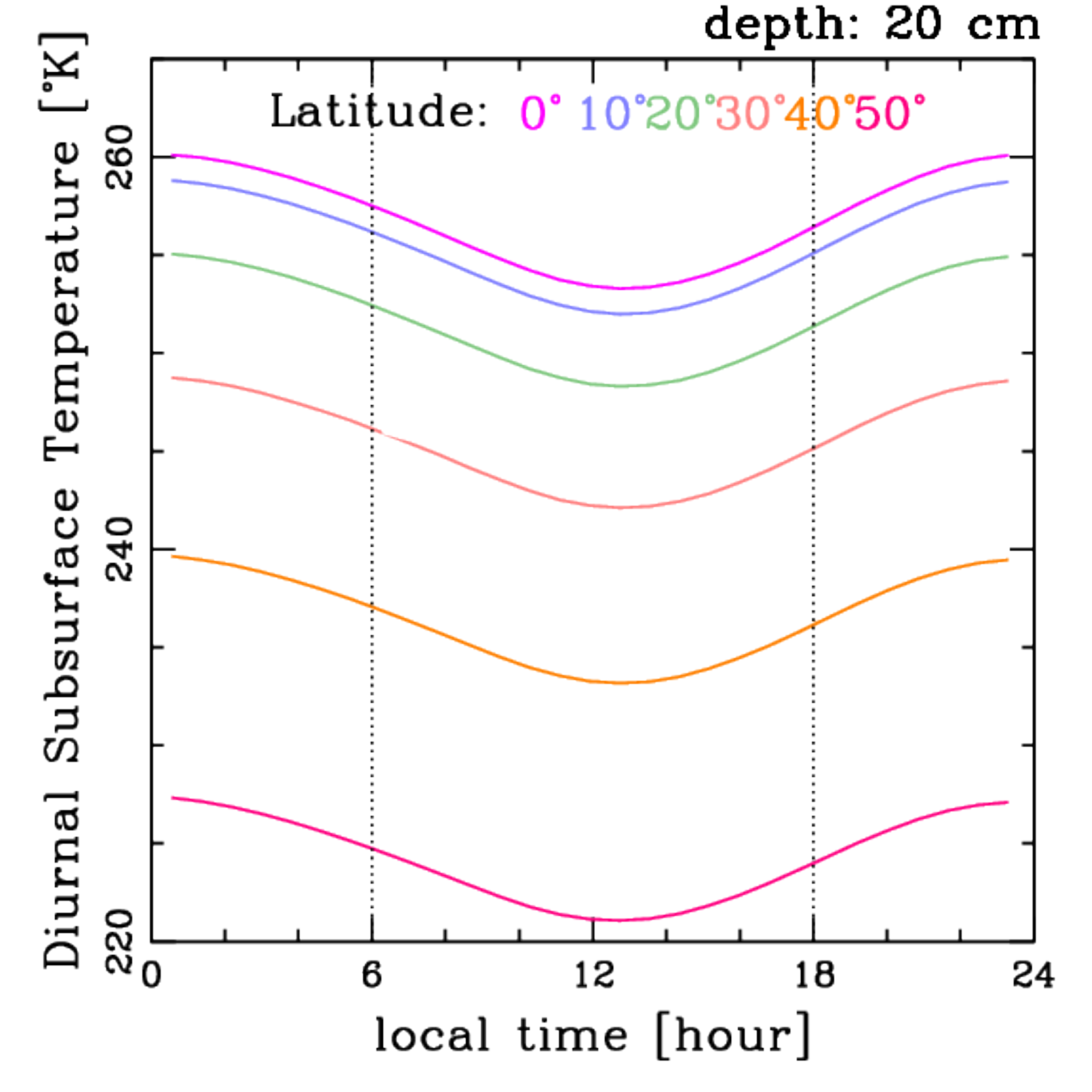}
\includegraphics[trim =0.0mm 5.0mm 0.0mm -8.0mm, clip,height=0.52\columnwidth]{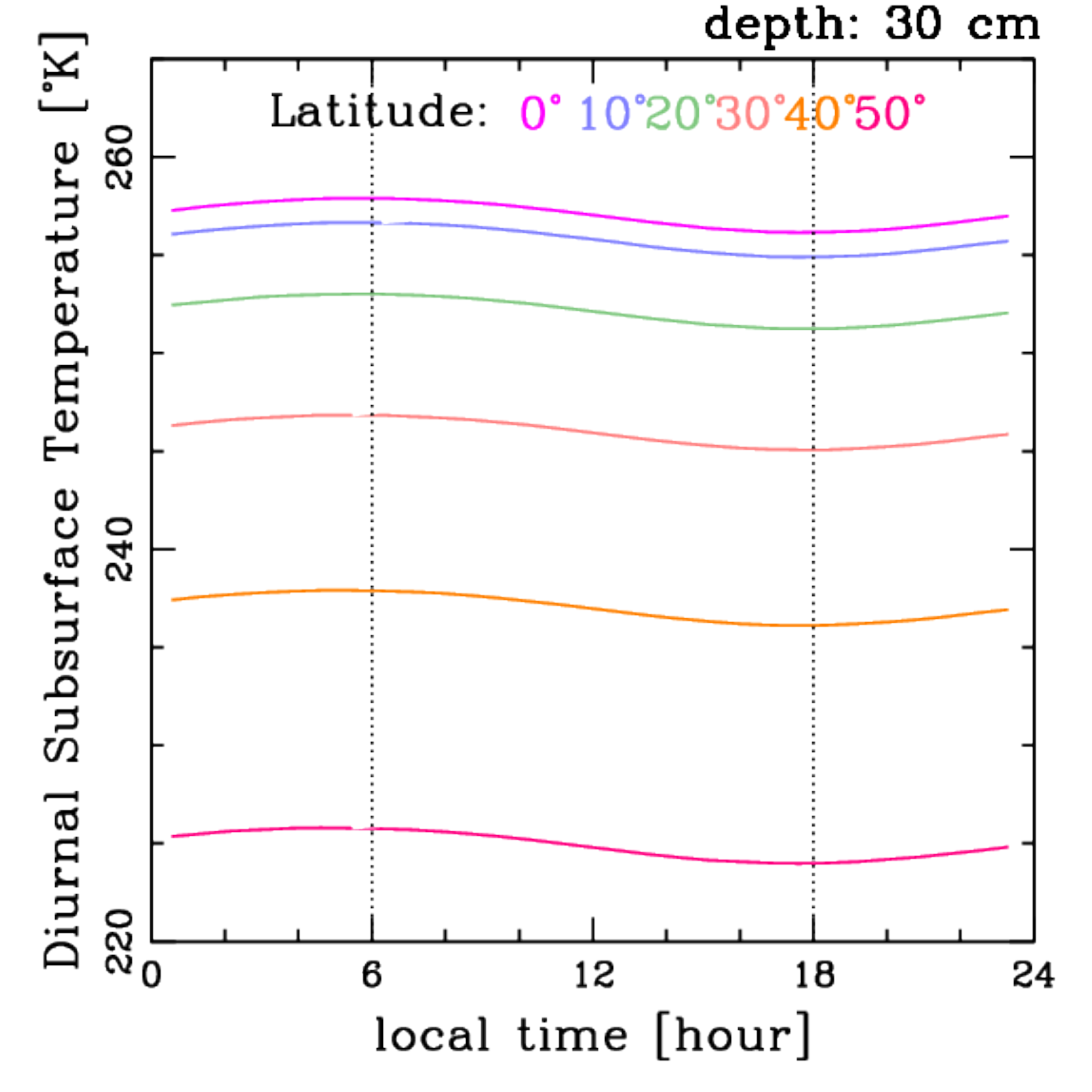}
\caption{
{\bf Upper left} Diviner Temperature measurements at different latitudes function of the lunar local time. {\bf From the left to the right, and from the top to the bottom} we show subsurface temperatures drawn from Diviner thermal models 
at 1, 2, 3, 4, 5, 10, 20 and 30 cm deep, respectively. These temperatures have been constrained 
by the Diviner observations using the methodology detailed in  \cite{Siegler:2011JGRE..116.3010S}.
}
\label{fig:surface_temp_model}
\end{center}
\end{figure*}

The epithermal-neutron counting rate and spacecraft altitude 
time series are shown in Figure~\ref{fig:lptimeseries}, along with other parameters (neutron sensor temperature and pulse height peak-channel position) used in this study.
All data shown in Figure~\ref{fig:lptimeseries}
are coloured based on the different mission phases described in Table 2 of \cite{Maurice:2004JGRE..109.7S04M}.  Green shows data from the High 1 phase (average altitude $\sim$100 km), cyan shows data from the High 2 phase (average altitude $\sim$100 km, flipped spacecraft spin axis), magenta shows data from the Low 1 phase (average altitude $\sim$40 km), and violet shows data from the Low 2 phase (average altitude $\sim$30 km).  
The origin of time was set at 1st of  January 1998 

\medskip Hereafter we will use data within the latitude range $[-55 \degree,$ $55 \degree]$, coming either from the 
High 1 (hereafter High Altitude) or Low 2 (hereafter, Low-altitude) phases, respectively. A third dataset with 
all the measurements within the same latitude range  will be termed Overall.  The number of 8s measurements within  Overall, High altitude and Low-altitude datasets are 2,851,632, 1,324,569 and 939,668, respectively.
With this choice of latitudes we are considering those parts of the lunar surface whose epithermal signal is not dominated by the hydrogen 
enhancements
 in the top few decimeters of regolith.  The individual count rates were normalized to altitude 30 km \citep{Maurice:2004JGRE..109.7S04M}.

\subsection{Lunar Prospector Instrumental Temperature and Energy spectrum information}
The LPNS included a temperature sensor. Its 
time-series variation is 
presented in panel {\bf c)} of Figure~\ref{fig:lptimeseries}. 
As with the count rate data we will use 
data within $|\mbox{latitude}|<55^{\degree}$.

The LPNS returned pulse height spectra of the $^3$He neutron captures, and a key feature is the 
764 keV full-energy capture peak (see Figure~\ref{fig:peak_temp}).
Panel {\bf d)} of Figure~\ref{fig:lptimeseries} shows the Cd $^3$He sensor peak-channel position versus time. 
The large jump in peak-channel position around time $\sim$300 days coincides 
with an increase  in the detector's high voltage. 

\subsection{Diviner/LRO temperature diurnal variation data}

The Diviner Lunar Radar Experiment on board LRO is an imaging multi-channel solar reflectance and infrared radiometer. This 
instrument maps the temperature of the lunar surface at $<$ 500 meter horizontal scales.
Diviner data sets are produced by the Diviner Science Team at the University of California, Los Angeles.

Here we use  temperature information as function of the lunar local time at different 
depths of the lunar subsurface. We will employ
the Diviner measurements and  model predictions constrained by such measurements 
\citep[see][for details]{Siegler:2011JGRE..116.3010S}.  Upper left panel in Figure~\ref{fig:surface_temp_model} shows the Diviner measurements and the remaining
eight
panels show  the model's predictions at depths of 1, 2, 3, 4, 5, 10, 20 and 30 cm. 

As expected,  subsurface temperatures have a  strong dependence on the local time, depth and latitude.  At the surface, temperature peaks at the local midday. At deeper depths, the temperature 
maximum shifts towards later times of the day, as expected for a diurnal temperature wave.

\section{Analysis}\label{sec:analysis}
\begin{figure}[ht]
\centering
\includegraphics[trim = 0.0mm 75.0mm 0.0mm 60.0mm, clip,width=0.95\columnwidth]{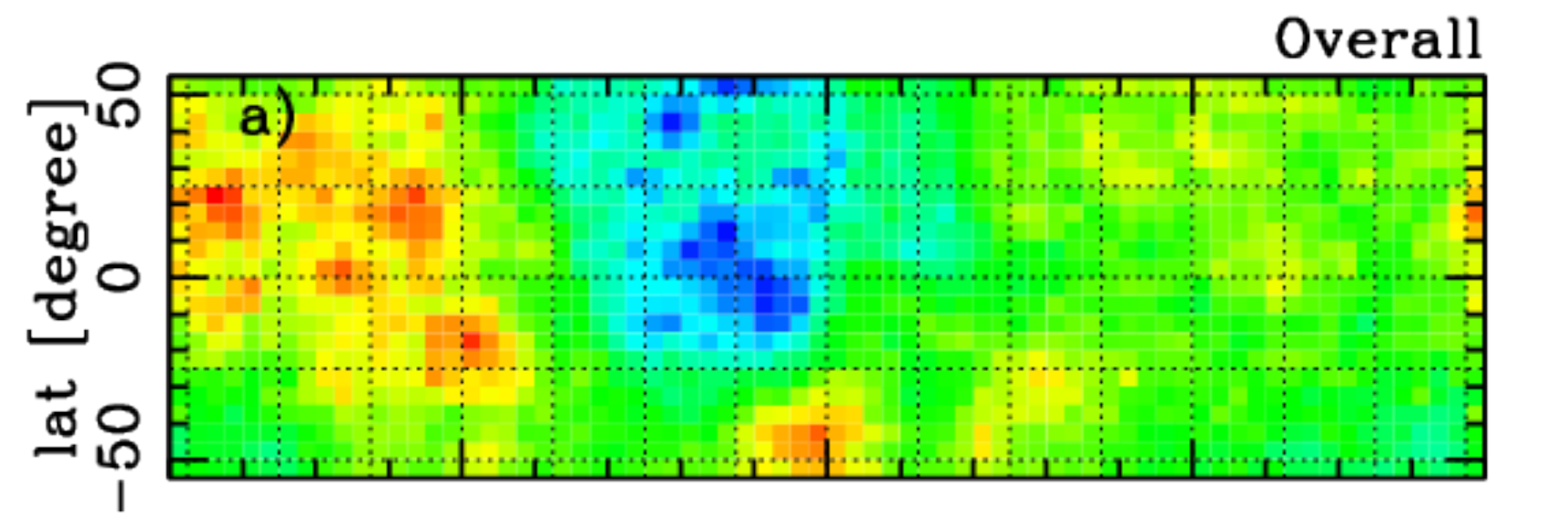}
\includegraphics[trim = 0.0mm 75.0mm 0.0mm 60.0mm, clip,width=0.95\columnwidth]{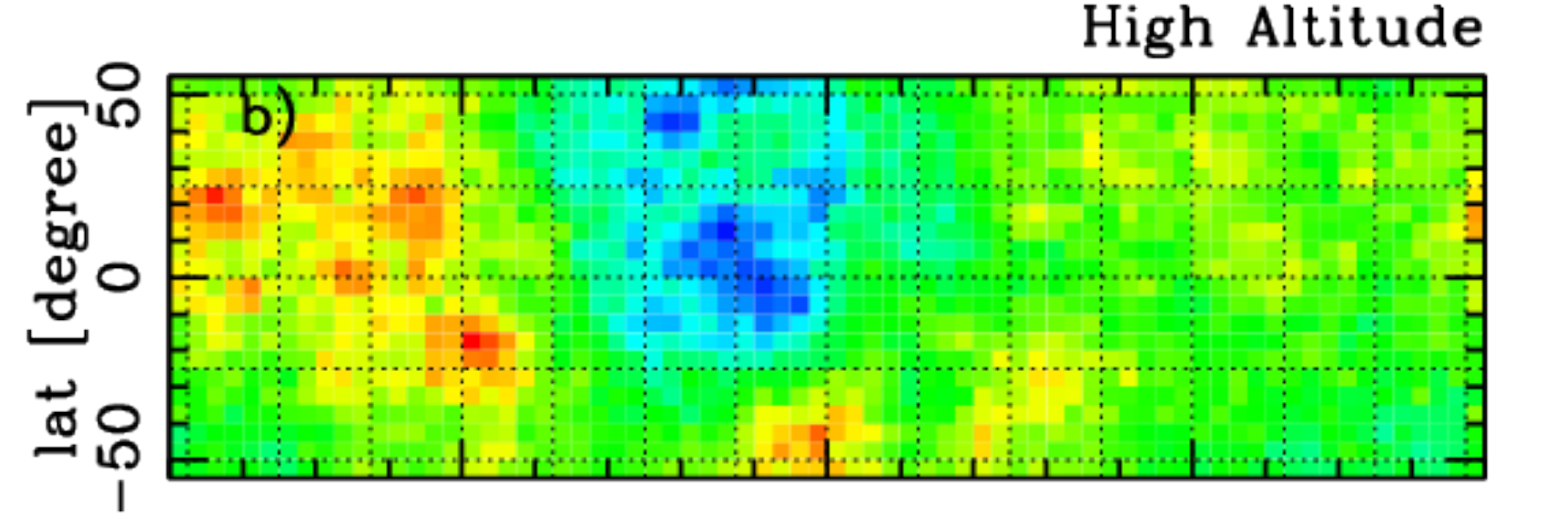}
\includegraphics[trim = 0.0mm 35.0mm 0.0mm 60.0mm, clip,width=0.95\columnwidth]{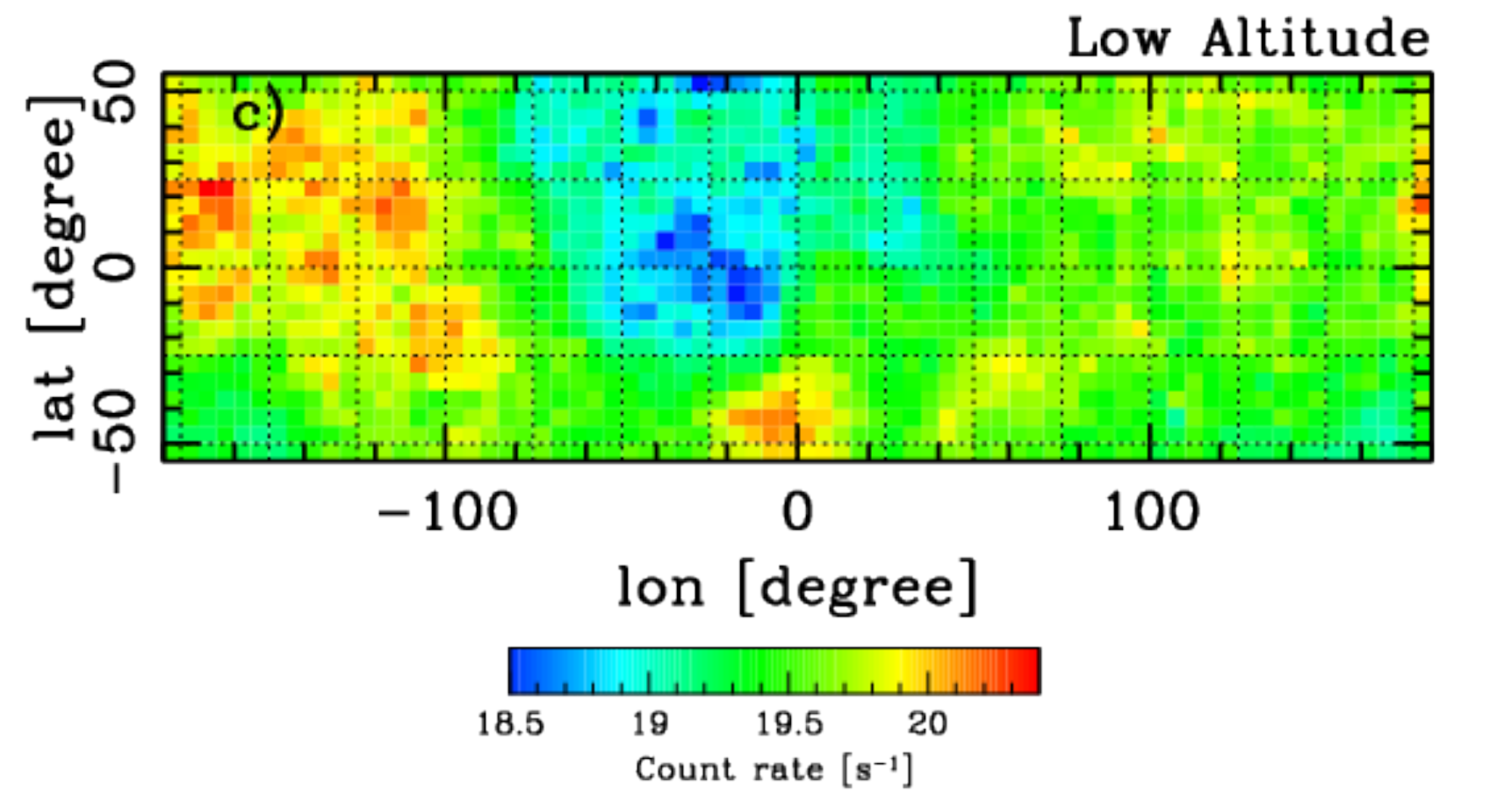}
\caption{Fiducial states adopted in this article. {\bf a)} Overall data. {\bf b)} High Altitude. 
{\bf c)} Low-Altitude. The data is pixellized in 5$\degree$$\times$ 5$\degree$ bins.} 
\label{fig:fiducial_map}
\end{figure}

\begin{figure}[h]
\begin{center}
\includegraphics[trim = 0.0mm 106.0mm 0.0mm 10.5mm, clip,width=0.90\columnwidth]{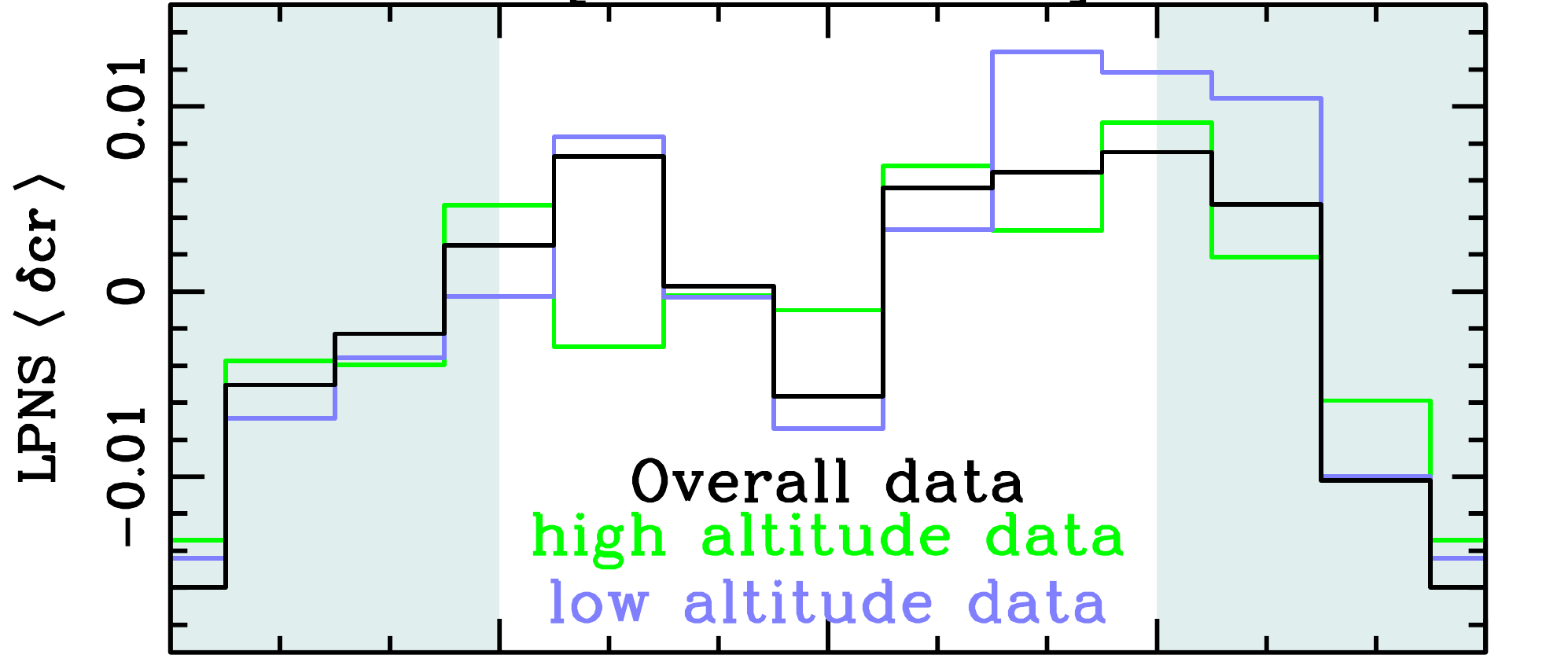}
\includegraphics[trim = 0.0mm 85.0mm 0.0mm 10.5mm, clip,width=0.90\columnwidth]{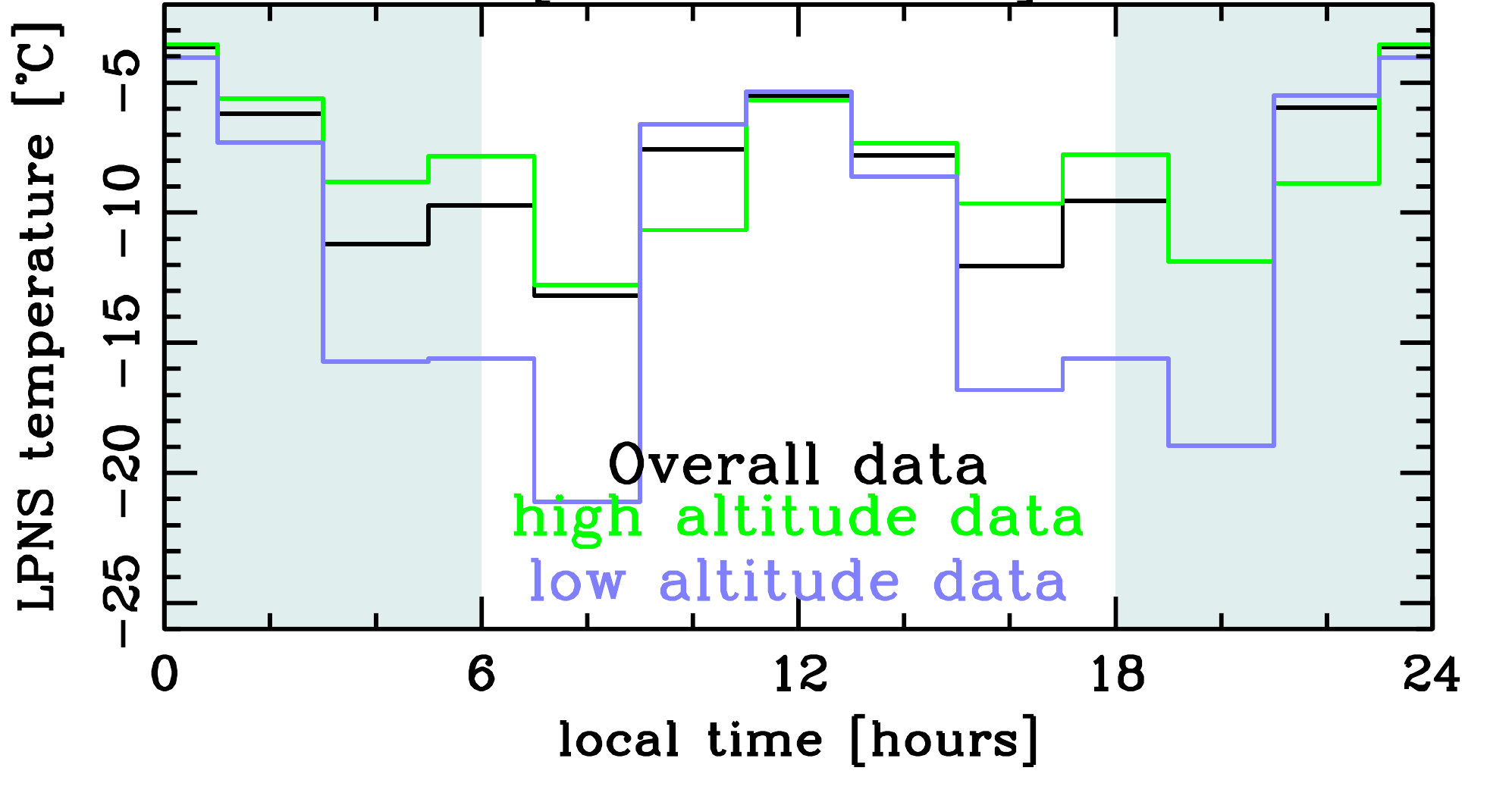}
\caption{
{\bf Upper panel}  Average {overcount rate} per {\it local time} bin. The {\it fiducial map} is 
the mean counting rate at a given location over all the time series.
The random errors in the measurements are smaller than 2.1$\times$10$^{-4}$,  5.2 $\times$10$^{-4}$, 3.5$\times$10$^{-4}$ for the overall (black), high altitude (green) and  low-altitude (violet) data, respectively. 
{\bf Lower panel} Average NS {temperature}  per { local time} bin. The random errors in the 
measurements are smaller than 3.4$\times$10$^{-2}$,  8.2 $\times$10$^{-2}$, 5.2$\times$10$^{-4}$ 
${\degree}C$ for the overall (black), high altitude (green) and  low-altitude (violet) data, respectively. 
The shaded regions of the diagrams represent the nightside.
The $x$-axis tickmarks coincide with the local time bin centers.
}
\label{fig:lp_deltaf_per_lt}
\end{center}
\end{figure}

To quantify the variability of the epithermal count rate, $\cre (\mathbf{x,}$ $\mathbf{ t})$,  
we introduce the stochastic {\it overcount rate} variable, $\delta \cre$:
\begin{equation}\label{eq:deltacr}
\delta \cre (\mathbf{x,t}) \equiv \frac{\cre(\mathbf{x,t})-\crf (\mathbf{x})} {\crf (\mathbf{x})} =\frac{\cre(\mathbf{x,t})} {\crf (\mathbf{x})}-1,
\end{equation}
where $\crf(\mathbf{x}) \equiv  \crf(\mbox{lon.},\mbox{lat.})$ is the fiducial count rate map
at a given local time instant.  Thus, the overcount rate field quantifies the 
change in counting rate,  $\cre(\mathbf{x,t})-\crf (\mathbf{x})$,  relative to a {\it fiducial state}, $\crf (\mathbf{x})$.  Hereafter, we will present results obtained using a fiducial state defined as  the mean epithermal counting rate in a two-hour bin centered around 6\,pm = +90 degrees from the sub-solar longitude on the Moon.  
Figure~\ref{fig:fiducial_map} presents the fiducial count rate maps for the overall, high altitude and low-altitude datasets.
$\cre(\mathbf{x,t})$ and  $\crf(\mathbf{x})$ are pixellated in 5$\degree$$\times$ 5$\degree$ bins: $\cre (\mathbf{x}_{ij},t_k)$ and  $\crf(\mathbf{x}_{ij})$, where  the spatial coordinate 
$\mathbf{x}$ is partitioned in latitude $i \in \{1, \cdots, 36\}$ and longitude $j \in \{1, \cdots, 72\}$, while the two-hour local time bins are represented by $k \in \{1, \cdots, 12\}$.  
In Figure \ref{fig:lp_deltaf_per_lt} we present the averaged overcount rate, $\langle \delta cr (\mathbf{x}_{ij},t_k)\rangle _{ij}$, in two-hour local time bins centered around $t_k$
\begin{equation}
\langle \delta cr (\mathbf{x}_{ij},t_k) \rangle _{ij} \equiv \frac{1}{n_{\mbox{\scriptsize{pixels}}}(t_k)}\sum_{ij}\delta cr (\mathbf{x},t_k) W(\mathbf{x}_{ij}),
\end{equation}
where $\langle  \cdots \rangle _{ij}$ denotes the average over 
all the pixels at a given instant $t_k$, the sum is over all the spatial pixels ($i$, $j$),
$W(\mathbf{x}_{ij})$ is a spatial ``mask" applied to the data as the means to 
exclude the polar regions:
\begin{equation}
W(\mathbf{x}_{ij}) =  \left \{ 
\begin{array}{ll}
1, &  \mbox{at  latitudes} \in [-55 \degree,  55 \degree], \\\\
 0, & \mbox{everywhere else,} 
\end{array}
\right .
\end{equation}
and 
\begin{equation}
{n_{\mbox{\scriptsize{pixels}}}(t_k)} \equiv \sum_{ij}   W(\mathbf{x}_{ij}).
\end{equation}

The three derived time series (overall, high and low-altitude) show significant peak-to-peak variations, $\sim0.0273$,  given that the random errors are smaller than $3.0\times10^{-4}$. 

\medskip In principal, these epithermal-neutron local-time variations may be due to variations in bulk hydrogen concentrations across the lunar surface, which is an idea suggested by \cite{Livengood:2014LPI....45.1507L}
to explain similar variations in LRO/LEND data.  However, a comparison of LPNS and LEND data argue against such an interpretation.  Specifically, even though the magnitude of the LPNS and LEND local-time variations are similar (2.7\% for LPNS versus ~2\% for LEND), their qualitative behavior is fundamentally different.  The LEND data show a single maxima (6 local time [LT]) and minima (15 LT), whereas the LPNS data show two maxima (7 LT and 15 LT) and minima (12 LT and 24 LT).  If there were  true local-time hydrogen variations that caused a 2 to 2.7\% variation in epithermal-neutron count rates, the measured local-time behavior should be similar in both datasets.  The fact that their local-time behavior is different suggests that other non-hydrogen systematic variations may be responsible for the observed local-time variations.  Because the two sets of measurements were made with different instruments on different spacecraft, such non-hydrogen variations need not be identical in both datasets.  Here, we investigate possible sources of non-hydrogen systematic variations in the LPNS data.  If we can show that such systematic effects can account for all or most of the local-time variations, then we can conclude that varying hydrogen concentrations are not the cause of these variations.  Below, we discuss two such systematic effects: variations in LPNS sensor temperature and variations in lunar surface and subsurface temperatures. 

\subsection{Epithermal-neutron count rate dependence on LPNS sensor temperature}
We first investigate if there is evidence that LPNS sensor temperatures might be related to systematic variations in the epithermal-neutron count rate that have not yet been removed in the prior data reduction procedures \citep{Maurice:2004JGRE..109.7S04M}. 
Specifically, we examine if the varying sensor temperatures are related to gain variations (or a multiplicative shift) in the neutron sensor pulse height spectra.  If such gain variations exist, we then ask if such variations are related to the ~2.7\% count rate variations.  We note that the variations in the high and low-altitude data are qualitatively similar, so unless otherwise noted, we hereafter only carry out analyses for the low-altitude LPNS data. 

\begin{figure}[!t]
\begin{center}
\includegraphics[trim = 0.0mm 00.0mm 0.0mm 00.0mm, clip,width=0.95\columnwidth]{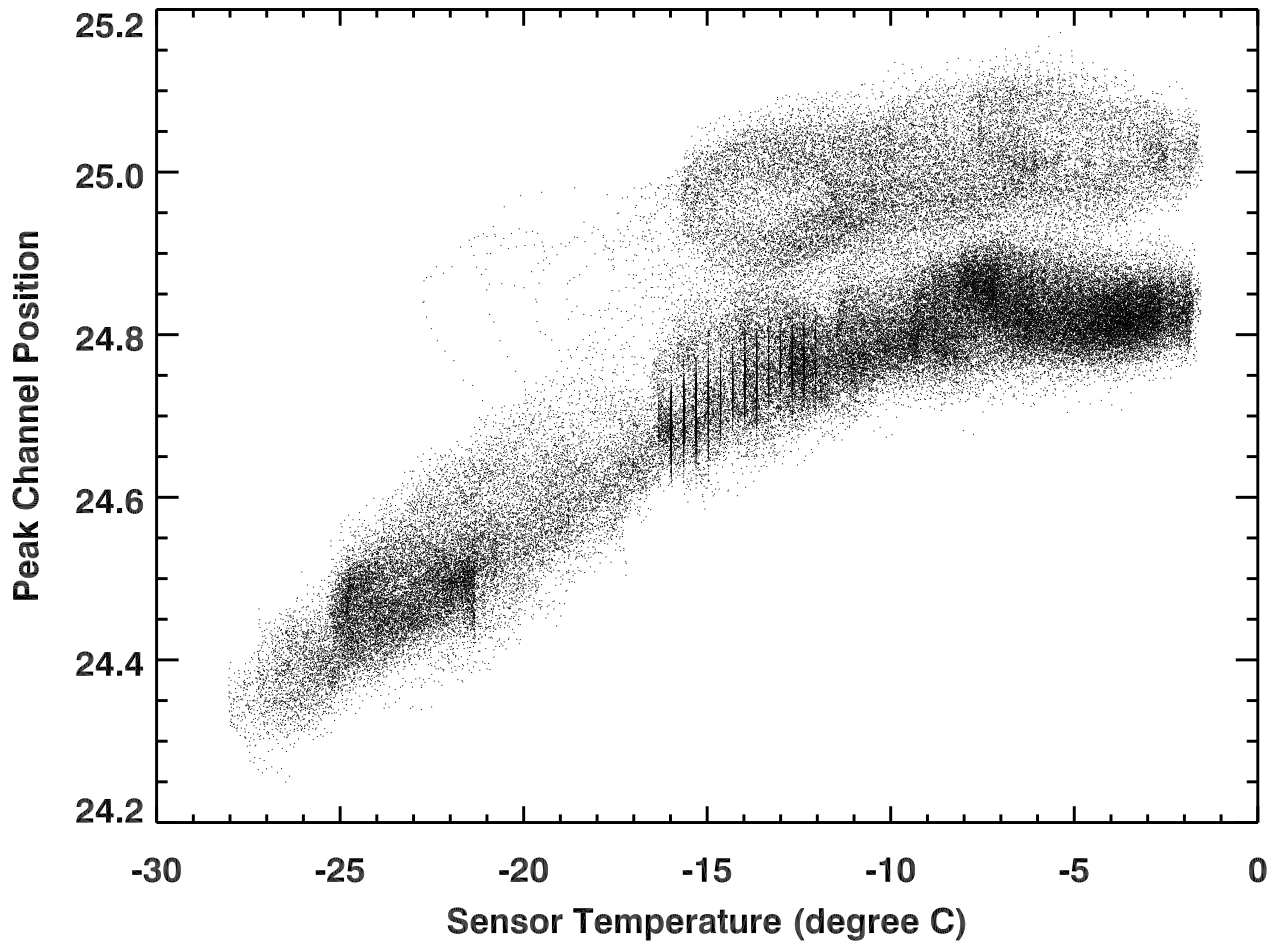}
\includegraphics[trim = 0.0mm 00.0mm 0.0mm 00.0mm, clip,width=0.95\columnwidth]{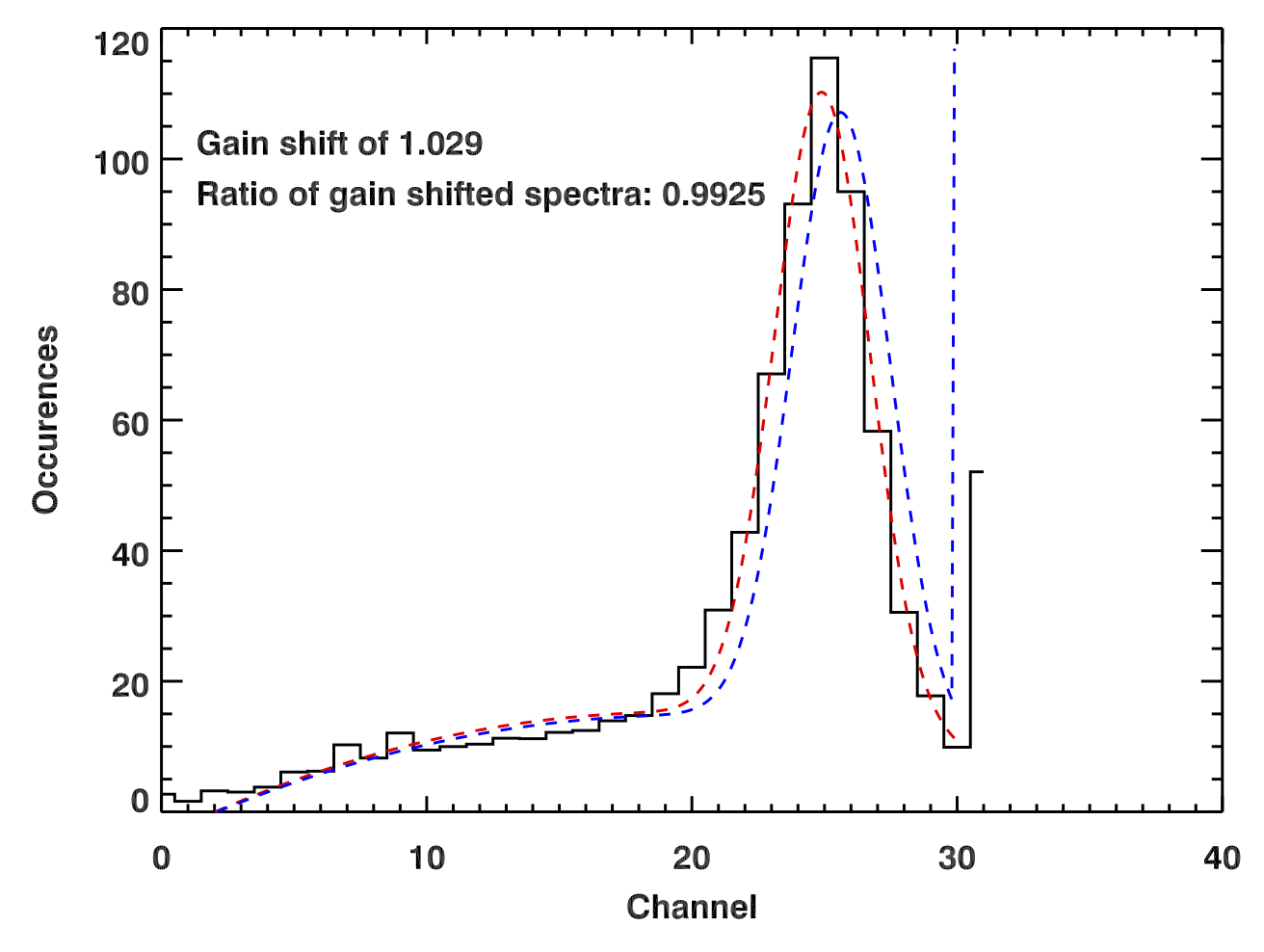}
\caption{{\bf Upper panel} Low-altitude {Peak-channel Position} versus { instrumental temperature}. {\bf Lower panel}. {Pulse height} vs { Channel}  from the LP Cd$^3$He sensor. Original and gain shifted spectra are denoted in dashed lines: original in red and gain shifted in blue (see text for details).
}
\label{fig:peak_temp}
\end{center}
\end{figure}

Inspection of Figures~\ref{fig:lptimeseries}c and \ref{fig:lptimeseries}d shows 
there is a likely
correlation between the sensor temperature 
and the gain. Figure~\ref{fig:peak_temp} (upper)
shows this correlation explicitly, where the temperature versus peak-channel position for the low-altitude data is
plotted.  
It is clear that the gain of the sensor shows a positive, but non-linear variation with sensor temperature. The dynamic range of the variation is about 0.5 to 0.7 channels, or a 
2 to 3\% variation.

To investigate if such a gain change could cause a count rate variation of 1 to 2\%, 
Figure~\ref{fig:peak_temp} (lower) shows a summed pulse height spectrum from the 
Cd $^3$He sensor. 
The LPNS pulse-height spectra were measured with 32 channels, 
so with such data it is difficult to see
if a 0.5 channel difference in gain could cause a counting rate shift. We therefore fit the spectra 
with a gaussian plus polynomial between channels 10 and 30 and plotted the function using 
300 channel values instead of 32 (red dashed line). To simulate the effect of a temperature 
dependent gain-shift, we shifted the spectra using the largest possible peak-position ratio allowed by the data (25.0/24.3). Both the original and gain shifted spectra
are shown in dashed lines (original in red; gain shifted in blue). When one sums the
counts between channels 15 and 30 (which are the channels used by \cite{Maurice:2004JGRE..109.7S04M}
for the count rate change measurements), there is a 0.7\% counting rate change. While this 
count rate change is
smaller than the measured change, it is in the same direction as the measured data, such that lower gain data (which
corresponds to lower instrument temperature) shows a larger counting rate, and therefore demonstrates that a varying gain caused by a varying temperature can result in varying LPNS epithermal-neutron count rates.

\subsection{Neutron Spectrometer sensor temperature correction}
\begin{figure}[!t]
\begin{center}
\includegraphics[trim = 0.0mm 45.0mm 0.0mm 0.0mm, clip,height=0.75\columnwidth]{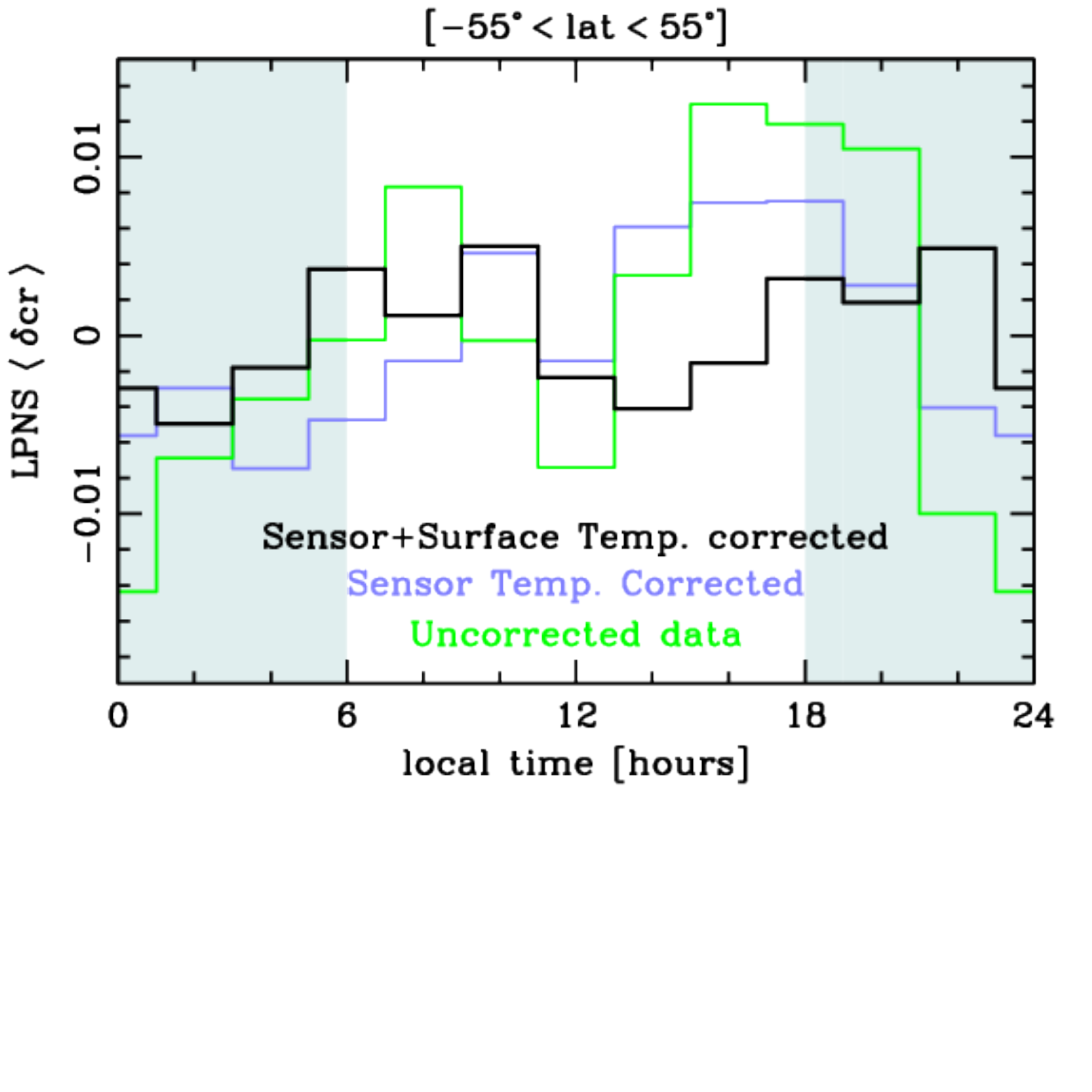}
\includegraphics[trim = 0.0mm 45.0mm 0.0mm 0.0mm, clip,height=0.75\columnwidth]
{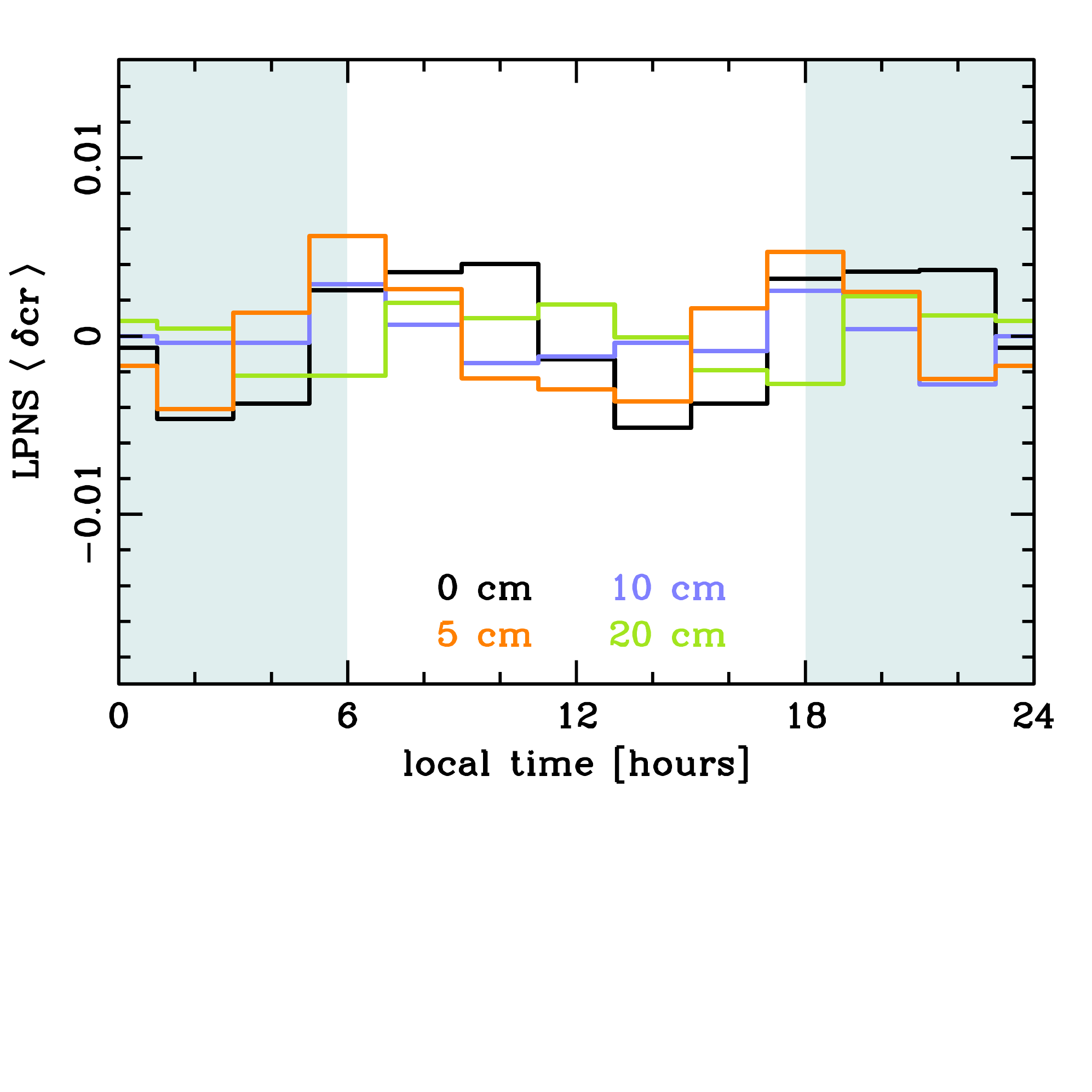}
\caption{Low-altitude count rate versus local time.  {\bf Upper panel}
Black  denotes the uncorrected count rate while 
violet and green represent sensor temperature corrected and sensor plus surface corrected, respectively. 
{\bf Lower panel} Subsurface temperature corrected count rates. Black, orange, violet and green represent count rate  data corrected with Diviner subsurface temperature models at 0, 5, 10 and 20 cm, respectively. The random errors in all the measurements are smaller than 3.4$\times$10$^{-4}$.
}
\label{fig:low_altitude_corrected_data}
\end{center}
\end{figure}

To remove the sensor temperature induced count rate variation, 
we have implemented the following two-step procedure: {\it i)} find the linear correlation coefficient between low-altitude time-series count rates and their associated sensor temperatures, and  {\it ii)}  use the linear relationship between those two variables to ``de-trend" the data. 
A least squares fit to the data produces 
\begin{equation}
cr =  -0.0194 \cdot T_{\scriptsize{sensor}}+19.170, 
\end{equation}
  where  the sensor temperature, $T_{\scriptsize{sensor}}$, is expressed in $^{\circ} C$ and the measured count rate, $cr$, in $s^{-1}$. 
  This relationship is applied to produce the sensor temperature-corrected count rates,
  \begin{equation}
  cr'\equiv cr + 0.0194 \cdot \left( T_{\scriptsize{sensor}}-\bar{T}_{\scriptsize{sensor}}\right),
  \end{equation}
where $\bar{T}_{\scriptsize{sensor}}$ denotes the mean sensor temperature.
These corrected count rates lead to fluctuations in the low-altitude overcount rates that are reduced from 2.73\% to 1.60\%, as shown in Table~3  and the violet line in the upper panel of Figure~\ref{fig:low_altitude_corrected_data}. In the same panel of Figure~\ref{fig:low_altitude_corrected_data} we also present the uncorrected data (green).

\begin{table}
\begin{center}
\begin{tabular}{lcc}
 \\ \hline \hline\\[-2pt]
 Soil type  & $a_{soil}$ & $b_{soil}$ \\
 \\[1pt]
 \hline
 \\[2pt]
FAN & 20.80 & 0.00280  \\[1.5pt]
Lunar 20 (L20)  & 20.70 & 0.00220 \\[1.5pt]
Apollo 11 (A11)  & 20.40 & 0.00160  \\[1.5pt]
Apollo 12 (A12)   & 20.08 & 0.00211 \\[1.5pt]
Lunar 24 (L24)  & 19.97 & 0.00163 \\[1.5pt]
\\ \hline \hline\\[-2pt]
\end{tabular}
\label{table:soil_parameters}
\end{center}
\vspace{-0.5cm}\caption{Various soil parameters,  $(a_{soil}, b_{soil})$, considered in this work. These parameters 
were obtained using a linear least squares fit to the points in panel a) of Figure~\ref{fig:countrates_vs_surftemp}. For further details regarding these soils see \cite{Lawrence:2006JGRE..11108001L}.}
\label{table:soil_parameters}
\end{table}

\subsection{Count rate correlation with lunar surface and subsurface temperature}
 \cite{Little:2003JGRE..108.5046L} and   \cite{Lawrence:2006JGRE..11108001L} used 
the Monte Carlo transport code MCNPX 
to study the transport of neutrons  in typical lunar soils and found that thermal and epithermal count rates are dependent on soil temperature.
Here, we consider a variety of subsurface temperatures as functions of both latitude
and local time, constrained by Diviner measurements \citep{Siegler:2011JGRE..116.3010S}. 

\begin{figure}[!t]
\begin{center}
\includegraphics[trim = 0.0mm 0.0mm 0.0mm 0.0mm, clip,width=0.98\columnwidth]{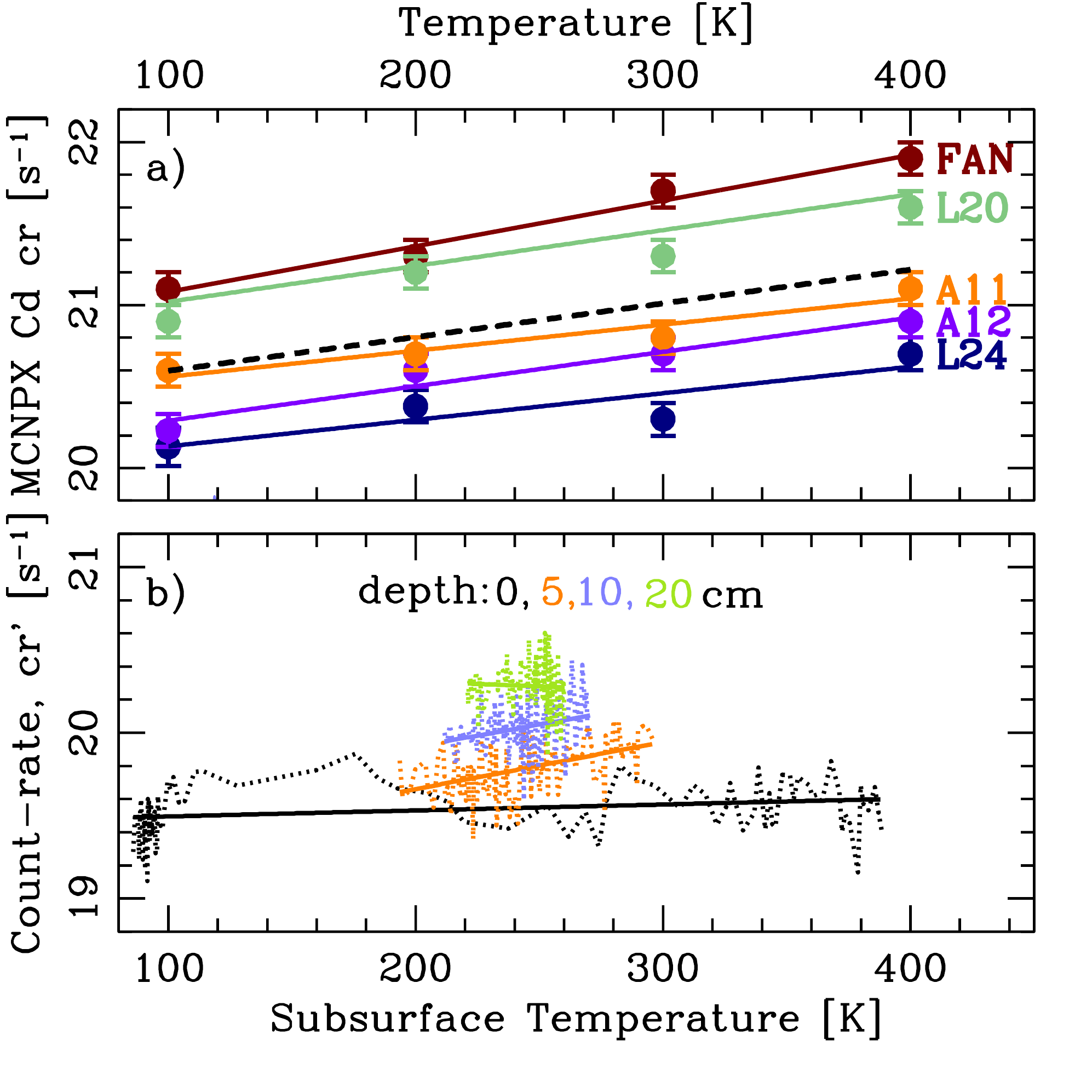}
\caption{ 
{\bf Panel a)} This panel, as in  \cite{Lawrence:2006JGRE..11108001L}, shows { 
count rates} versus {surface temperatures}.  Each color denotes a different soil:
FAN (brown), Lunar 20 (L20, powder green), Apollo 11 (A11, orange), Apollo (A12, violet) and Lunar 24 (L24, blue navy). The dashed black line denotes the ``average lunar soil".
{\bf Panel b)} The measured {count rates} versus { surface and sub-subsurface temperature}. 
Dotted lines represent the box-smoothed count rates. Solid lines denote the least  squares fits to the smoothed data. 
For clarity the 5, 10 and 20 cm curves are shifted upwards by 0.25, 0.50 and 0.75 $s^{-1}$, respectively.
}
\label{fig:countrates_vs_surftemp}
\end{center}
\end{figure}

\medskip To take into account the effects of the lunar subsurface temperature we consider an ``average soil"
made up of equal fractions of FAN, Luna 20 (L20), Luna 24 (L24), Apollo 11 (A11)  and Apollo 12 (A12). 
In panel a) of Figure~\ref{fig:countrates_vs_surftemp} we reproduce some of the 
 \cite{Lawrence:2006JGRE..11108001L}
results for epithermal-neutron count rates from these lunar soils.
The dashed black line represents an ``average lunar soil", 
with epithermal count rates having a  temperature dependence  given by
\begin{equation}
cr \equiv \bar{a}_{soil} + \bar{b}_{soil}  T_{soil}= 20.39+ 2.068 \times 10^{-2}  T_{soil},
\label{eq:average_soil}
\end{equation}
where the soil temperature $T_{soil}$ is expressed in Kelvin. 
In Table \ref{table:soil_parameters} we present the various soil parameters, $ (a_{soil}, b_{soil})$,   that are used in the calculation of $\bar{a}_{soil}$  and $\bar{b}_{soil}$ of 
Equation \ref{eq:average_soil}. These parameters result from linear least squares fits to the points in panel a) of Figure~\ref{fig:countrates_vs_surftemp}.

\medskip (Sub)surface temperatures are derived using the Diviner 
measurements as well as the models constrained by these measurements shown in Figure~\ref{fig:surface_temp_model}. To obtain the proper temperatures for an 
epithermal-neutron temperature correction,
we interpolate in local time and latitude at a given depth. 
\begin{table}[!t]
\begin{center}
\begin{tabular}{lcr}
 \\ \hline \hline\\[-2pt]
Depth  & $a'_{depth}$ & $b'_{depth}$~~~ \\
 \\[1pt]
 \hline
 \\[2pt]
0 cm  & 19.46 & 0.00036  \\[1.5pt]
1 cm   & 19.38 & 0.00067 \\[1.5pt]
2 cm  & 19.19 & 0.00143 \\[1.5pt]
3 cm  & 18.98 & 0.00227 \\[1.5pt]
4 cm  & 18.98 & 0.00227 \\[1.5pt]
5 cm  & 18.84 & 0.00284 \\[1.5pt]
10 cm  & 18.88 & 0.00266 \\[1.5pt]
20 cm  & 19.67 & -0.00056 \\[1.5pt]
30 cm  & 19.46 & 0.00036 \\[1.5pt]
 \\ \hline \hline\\[-2pt]
\end{tabular}
\end{center}
\vspace{-0.5cm}\caption{Equation \ref{eq:counts_temperature_depths}
linear least squares fit parameters,  $(a'_{depth}, b'_{depth})$, at the various depths used to model the epithermal count rates versus soil temperature.}
\label{table:depth_parameters}
\end{table}
We  fit a linear curve to the box-smoothed count rates versus (sub)surface temperature at various depths, such that
\begin{equation}
cr'_{depth}=a'_{depth}+b'_{depth} T_{soil,\,depth},
\label{eq:counts_temperature_depths}
\end{equation}
where $ T_{{{soil,\,depth}}}$ is expressed in Kelvin. 
The values of $a'_{depth}$  and $b'_{depth}$ for the various depths are shown in Table~\ref{table:depth_parameters}. 
In panel b) of Figure~\ref{fig:countrates_vs_surftemp} we show the count rates corrected for instrumental temperature, $cr'$,
versus  (sub-)surface temperatures at various depths.
The count rates are box-smoothed using the same number of observations per bin and are denoted by dotted lines. The corresponding solid lines  
represent the linear least squares fits at each depth. For the sake of the temperature corrections, the given depth represents the ``effective" depth from which the neutrons are emitted and is an approximation to the actual scenario where detected neutrons are emitted from a distribution of depths.  

\medskip To match the ``average lunar soil" count rate versus soil temperature predictions shown at the upper panel of Figure~\ref{fig:countrates_vs_surftemp}
and the LPNS-Diviner measurements of the same relationship shown at the lower panel of the same figure we introduce the following transform:
\small
\begin{eqnarray}
 cr_{final,\,depth}&=&cr'_{depth}+\nonumber \\
  &  &\left(\bar{a}_{soil}-a'_{depth}\right)+\bar{T}_{soil,\,depth}\left(b'_{depth}-\bar{b}_ {soil}\right) +\nonumber \\
& &\left(\bar{b}_{soil}-b'_{depth}\right)\left(  T_{soil,\,depth}-\bar{T}_{soil,\,depth}\right ). 
\label{eq:fitting_fits}
\end{eqnarray}
\normalsize
Here, $\bar{T}_{soil,\,depth}$ is the average soil temperature at a given depth. 
Briefly, the transform is made up of two parts: 
{\it i)} move the measured point linear upwards is such a way that makes it coincide with the value of the model predictions at  $\bar{T}_{soil,\,depth}$ (second and third terms of the rhs in the above equation), and 
{\it ii)} make the slope of the linear fit to the measured points match the model 
counterpart, which is denoted by the last term in Equation~\ref{eq:fitting_fits}.

In the upper panel of
Figure~\ref{fig:low_altitude_corrected_data}
 we show the average overcount rate, $\langle \delta cr \rangle$, corrected
for sensor temperature and surface temperature
effects (black). In the lower panel of the
same figure we show the overcount rate as corrected
for subsurface temperatures at different
depths. Table~3 
contains the peak-to-peak
variation of the overcount rate for different subsurface
temperature corrections, with the minimum
variation occurring for the subsurface 
depth of 20 cm.  This result shows that if we correct the LPNS epithermal-neutron 
count rates for temperature at an effective depth of 20 cm, we can minimize the local-time variations.  Because the remaining variation is quite small ($<$0.5\%), we therefore find that there is no need to invoke local-time varying hydrogen concentrations to explain most of the LPNS local-time epithermal-neutron variations.  We further note that 
\cite{Little:2003JGRE..108.5046L}
predicted that the ``effective" depth for thermal-neutron emission is 30 g/cm$^2$, which for a soil density of 1.5 g/cm$^3$, results in a physical depth of 20 cm.  This is the same value as our data-constrained effective depth for epithermal-neutrons, and supports the interpretation that much of the measured local-time neutron variations are caused by subsurface temperature variations.

\begin{table}[!t]
\begin{center}
\tabcolsep 2.5pt
      \begin{tabular}{lccc}
   \\ \hline \hline\\[-2pt]
   {LPNS } &~~~~{ }~~~~ & ~~~~{ }~~~&
 ~~~{ }  
   \\[2pt]
 {sub-dataset} &
 ~~~~{$\left < \delta cr\right >_{\scriptsize{min}}$}~~~~ & 
 ~~~~{$\left < \delta cr\right >_{\scriptsize{max}}$}~~~& 
 ~~$\Delta cr$  \\[2pt]\\ 
\\   \hline \\
Overall  &  -0.0160  &  0.0075 & 0.0235 \\[1pt]
High altitude  &  -0.0134  & 0.0091  &  0.0226\\[1pt]
Low-altitude &    -0.0144  & 0.0130  &  0.0273\\[3pt]
\hline 
 \hline \\
\small{Low-altitude}\\
\small{sensor temperature}\\[1pt]
\small{corrected} &   -0.0075  &  0.0075   & 0.0150 \\[3pt]
\hline 
\hline \\[3pt]
 Low-altitude \\
 subsurface \\
 temperature\\
 corrected \\
\hline \\
 Depth 0 cm  &   -0.0052 & 0.0040 &  0.0092\\[1pt]
 Depth 1 cm  &   -0.0044& 0.0048  &  0.0092\\[1pt]
 Depth 2 cm  &   -0.0053 & 0.0044 &  0.0098\\[1pt]
 Depth 3 cm  &   -0.0056 & 0.0052&  0.0108\\[1pt]
 Depth 4 cm  &   -0.0056 & 0.0052&  0.0108 \\[1pt]
 Depth 5 cm  &   -0.0042 & 0.0057  &  0.0098\\[1pt]
 Depth 10 cm  &   -0.0028 & 0.0029 &  0.0057\\[1pt]
 Depth 20 cm  &   -0.0027 & 0.0022  &  0.0049\\[1pt]
Depth 30 cm & -0.0052 & 0.0040 &  0.0092\\[1pt]
      \hline \hline 
      \footnotetext{first semestre}  
    \end{tabular}
    \normalsize
     \caption{ Peak-to-peak variation of the LPNS overcount rate variation, $\Delta cr \equiv \langle  \delta cr \rangle_{\scriptsize{min}}-\langle  \delta cr \rangle_{\scriptsize{min}}$, of the different datasets applied in this article. 
}   
\end{center}
\label{table:peak_to_peak}
\end{table}

\section{Discussion and Conclusions}\label{sec:conclusions}

In the prior section, we have shown that the LPNS epithermal-neutron count rate data equatorward of 55$^\circ$ show a local-time variation that has a magnitude of 2.7\%.  We have found that corrections for varying LPNS sensor temperatures and lunar subsurface temperatures can reduce this 2.7\% variation to less than 0.5\%.   \cite{Livengood:2014LPI....45.1507L} have also found $\sim$2\% local-time variations in the LEND epithermal-neutron count rates.  Their interpretation of this result was that these variations were cause by mobile hydrogen that deposits once the sun sets, since the surficial temperature is low enough to lead to a long residence time during the night.  It then leaves the lunar surface at dawn as higher temperatures imply very short residence times 
\citep[see][]{Schorghofer:2007JGRE..112.2010S}
for a definition of residence time and its dependence on temperature).  Because we can quantitatively account for all but 0.5\% of the LPNS local-time variation, we find 
little-to-no support for the interpretation that varying hydrogen concentrations are the dominant cause of the local-time epithermal-neutron variations in either the LPNS or LEND datasets.  

Even though the remaining LPNS local-time variations do not exhibit the time dependence expected for diurnal variations suggested by Livengood et al. (2014), we can nevertheless investigate the implications if the remaining 0.5\% local-time variations are due to 
time-variable hydrogen concentrations.  Specifically, if a 0.5\% diurnal change in epithermal-neutron counting rate were the result of mobile water molecules, then how much water would this imply? We note that for this type of diurnally varying water, the water stratigraphy would likely take the form of a thin wet layer over a thicker dry layer.  In this case, we use results from \cite{Lawrence:2011JGRE..116.1002L} who modeled epithermal-neutron count rates for wet-over-dry layering and found that for top layers thinner than 5 to 10 cm, increasing amounts of hydrogen would result in a epithermal-neutron count rate increase rather than decrease.  Here, we therefore assume that hydrogen-bearing compounds are in an upper layer of limited thickness that participates in the diurnal condense/release cycle. We consider Figure~3 of  \cite{Lawrence:2011JGRE..116.1002L} and posit an upper, exchangeable regolith layer thickness of 3 g/cm2. The epithermal 
count rate in LPNS would increase by $\sim$0.5\% if that upper layer contained $\sim$1 wt\% water-equivalent hydrogen, i.e., 0.03 g/cm$^2$ of H$_2$O, over dry regolith. This 30 mg/cm$^2$ of exchangeable H$_2$O corresponds to 0.03/18 moles/cm$^2$ of H$_2$O molecules, or 10$^{21}$ molecules/cm$^2$. If exchangeable and mobile, this number is the column density of exospheric water that would be present over the dayside. Assuming a rough dayside scale height of 100 km (10$^7$ cm), the near-surface density of H$_2$O would be 10$^{14}$ molecules/cm$^3$. Preliminary reports show that observed levels of H$_2$O and OH are at least 10 orders of magnitude below this value \citep{Benna:agu2014}).  Other mobility scenarios that do not involve the lunar exosphere seem highly implausible.

Alternatively, we suggest that the remaining 0.5\% variation is likely to due other systematic variations that have not been fully taken into account.  As one suggestion, we consider the fact that the subsurface temperature correction approximated a distribution of neutron emission depths using a single ``effective" neutron emission depth.  While calculating the true distribution of neutron emission depths is beyond the scope of this current study, we can carry out a slightly more complicated (and possibly realistic) subsurface temperature correction by assuming that the neutrons are uniformly emitted throughout the top 30 cm.  With this assumption, we can reduce the prior 0.49\% local-time variation to less than 0.25\%.  While this final result should be treated with some caution because a true depth-dependent emission function has not been derived, it does demonstrate that the remaining 0.49\% variation shown in Table~3 can likely be reduced further without appealing to varying hydrogen concentrations.  

In summary, we have shown that the LPNS epithermal-neutron data do exhibit local-time variations that have a magnitude of 2.7\%.  However, these variations can be understood as being due to systematic variations of the LPNS sensor temperature as well as lunar subsurface temperatures.  An additional result of this work is that by showing the LPNS local-time variations can be minimized by a subsurface temperature at a depth of 20 cm, the LP neutron data have shown to be consistent with modeled subsurface temperatures, neutron-transport modeled temperature dependencies, as well as the predicted effective depth of subsurface neutron emission.

\vspace{0.1in}
\noindent{\bf{Acknowledgments}}
\newline
This work was primarily supported by a grant from the NASA Lunar Advanced Science and Exploration Research program to JHU/APL (NNX13AJ61G), and also received support from the NASA Solar System Exploration Research Virtual Institute.
VRE was supported by the Science and Technology Facilities Council (UK) [ST/L00075X/1].

\vspace{0.1in}

\noindent{{\bf References}}


\pagestyle{empty}

\end{document}